\documentclass[pra,twocolumn,showpacs,showemails,superscriptaddress,preprintnumbers,amsmath,amssymb]{revtex4-1}
\usepackage{amssymb}
\usepackage{amsfonts}
\usepackage{amsmath}
\DeclareMathOperator{\Tr}{Tr}

\usepackage{graphicx}
\usepackage{subfigure}
\usepackage{hyperref}
\begin{document}
\title{The Fidelity of Measurement-Based Quantum Computation under a Boson Environment}
\author{Jian Wang}
\affiliation{Peking University, Beijing 100871, China}

\author{Ding Zhong}
\affiliation{Peking University, Beijing 100871, China}

\author{Liangzhu Mu}
\email{muliangzhu@pku.edu.cn}
\affiliation{Peking University, Beijing 100871, China}

\author{Heng Fan}
\email{hfan@iphy.ac.cn}
\affiliation{Institute of Physics, Chinese Academy of Sciences, Beijing 100190, China}

\date{\today}

\begin{abstract}
We investigate the fidelity of the measurement-based quantum computation (MBQC) when it is coupled with boson environment, by measuring cluster state fidelity and gate fidelity. Two different schemes of cluster state preparation are studied. In the Controlled-Z (CZ) creation scheme, cluster states are prepared by entangling all qubits in $|+\rangle$ state with CZ gates on all neighboring sites. The fidelity shows an oscillation pattern over time evolution. The influence of environment temperature is evaluated, and suggestions are given to enhance the performance of MBQC realized in this way.
In the Hamiltonian creation scheme, cluster states are made by cooling a system with cluster Hamiltonians, of which ground states are cluster states.
The fidelity sudden drop phenomenon is discovered. When the coupling is below a threshold, MBQC systems are highly robust against the noise. Our main environment model is the one with a single collective bosonic mode.
\end{abstract}

\pacs{03.67.Pp, 03.67.Lx}

\maketitle

\section{Introduction}
Measurement-based quantum computation (MBQC) is a widely accepted scheme for quantum computation \cite{Raussendorf2001, Raussendorf2003, Briegel2009}. Instead of designing complicated quantum gates to manipulate qubits, MBQC is implemented by executing a sequence of single-qubit measurements on cluster states consisting of a group of highly entangled qubits.
As a result, one great difficulty lies in the cluster-state preparation.

There are various proposals to prepare cluster states. In optics, people use fusion operation to bind small cluster states into a larger cluster state \cite{Browne2005, Nielsen2004}. In quantum dots, people also know method to do it \cite{Guo2007}.
In Ref. \cite{Raussendorf2003}, Raussendorf \emph{et al.} pointed out two general ways to prepare cluster states.
The first one is to prepare all qubits in $|+\rangle$ state, and entangle them into a cluster state by implementing
CZ gates on all neighboring sites. The second one is to design a so-called cluster Hamiltonian, of which the ground state is a cluster state, and then cool down the system to obtain an approximate cluster state.
The idea of cluster Hamiltonian has been further explored. For example, one can encode four physical qubits into one logical qubit to achieve an experimentally realizable cluster Hamiltonian \cite{Bartlett2006}.
It is also shown that the topologically protected MBQC can reduce the thermal fluctuations \cite{Li2011} in the Hamiltonian-created cluster state.
Experiments on optical systems have been performed to demonstrate various quantum algorithms and protocols using the MBQC scheme. In 2005, Walther \emph{et al.} reported the demonstrative experiment on 4-qubit cluster states \cite{Walther2005}. In 2007, Grover's search algorithm on four qubits is performed \cite{Chen2007}.
Also, Deutsch's algorithm is realized by the MBQC on a four-qubit optical system \cite{Tame2007}. Remarkably, in the same year, six-photon cluster state was successfully entangled by Pan's Group \cite{Lu2007}.
The one-way MBQC scheme is even used to test the quantum version of the prisoner's dilemma \cite{Prevedel2007}.
A four-photon cluster state with very high fidelity is demonstrated in Ref. \cite{Tokunaga2008}.
The realization of MBQC beyond cluster state is also reported \cite{Gao2011}.

Unfortunately, cluster states, as a highly entangled system, are fragile to decoherence. It is thus important to analysis the noise to ensure the computation on a cluster state reliable. Some works have been done on this topic. In 2006, a method is proposed to check the fidelity of a four-qubit cluster state experimentally \cite{Tokunaga2006}.
The entanglement sudden death phenomenon \cite{yuting2009}, which may affect the fidelity of cluster states, is also studied \cite{weinstein2009}.
Recently, Fujii \emph{et al.} studied the error appeared in the Hamiltonian-created cluster states when the temperature is non-zero \cite{Fujii2013}. They discovered that the fidelity shows a sudden change at a certain threshold temperature.

\begin{figure}[!htbp]
  \centering
  \includegraphics[width=0.3\textwidth]{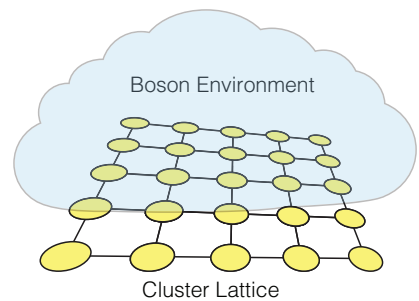}\\
  \caption{(Color online) A cluster state coupled with a boson environment.}\label{bosonenvironment}
\end{figure}

In this paper, we analyze the performance of the MBQC system when coupled with boson environment. The cluster state fidelity and four kinds of gate fidelity is measured, and the gate fidelity is studied in detail.

Due to the threshold theorem of fault-tolerant quantum computation, if the error in individual quantum gates is below a certain threshold, quantum computation in a large scale can be achieved as well \cite{qcqi}. As a result, our analysis works to protect MBQC systems with arbitrary scales.

Boson environment has long been a concerned issue in various topics \cite{Jouzdani2013}. More importantly, the boson environment, which is the noise caused by harmonic oscillators, actually describes a wide range of weak noises. Thus, this noise model is generic to many quantum computation cases.

A large cluster state can be prepared bit by bit. Therefore, a strategy against noise is to prepare a bit of a cluster state right before it is measured. In this paper, we assume that the cluster state for an individual gate operation is prepared at one time. With individual gate fidelities being analyzed, 
the fidelity of the whole MBQC can then be studied by the scheme of fault-tolerant quantum computation.

Two preparation schemes for the cluster state are evaluated. The first scheme is proposed by Briegel and Raussendorf \cite{Briegel2001}. In this scheme, cluster states are prepared by entangling all qubits previously in $|+\rangle$ state with Controlled-Z (CZ) gates on neighboring sites. The second scheme is proposed by Raussendorf later \cite{Raussendorf2003}, where cluster states are made by cooling a system with cluster Hamiltonians, of which cluster states are ground state. Both preparation is significant and widely applicable. However, since the two preparation schemes are quite different, it may not be meaningful to compare the fidelity between them.

This paper is organized as follows. In Sec. \ref{introToGF}, we introduce gate fidelity. In Sec. \ref{gatecreation}, we analyze how the coupling of a boson environment affects the cluster state entangled by CZ gates. We first solve the pure phase noise case exactly, and analyze this case in detail. Suggestions are given to minimize the damage caused by the coupling. Then, we consider both phase noise and amplitude noise, which is a more general case. We solve this problem numerically. In Sec. \ref{hamiltoniancreation}, we analyze the influence of a boson environment to the cluster Hamiltonian situation, with both phase and amplitude noise considered We also discover a threshold coupling coefficient, at which the fidelity drops dramatically. Sec. \ref{discussions} presents the discussion. We address the difference between gate fidelity and the corresponding cluster state fidelity. The collective character of our noise model is also discussed. We present our conclusion in Sec. \ref{conclusion}.

\section{Introduction to the Fidelity for Gate Operations}\label{introToGF}
The fidelity for a cluster state \cite{Raussendorf2003} is defined straightforwardly: 
\begin{equation}
F = \Tr \left(|\Psi_C\rangle\langle\Psi_C|\rho\right),
\end{equation}
where $|\Psi_C\rangle$ is a perfect cluster state, and $\rho$ is the state being judged. The form and utility of a cluster state $|\Psi_C\rangle$ can be found in Ref. \cite{Raussendorf2003}.

However, we are often more concerned about how good a gate operation is implemented by a MBQC system, and the cluster state fidelity fails to answer the question. It is thus required to define fidelity for gate operations. A good definition employs a process called gate teleportation.

The basic idea of gate teleportation is simple: if you apply some unitary operations to an EPR pair and use the pair to teleport a qubit, rather than getting the original information from the transported qubit, you will receive transformed information. By applying the proper gate to the EPR pair, we can get the transported qubit transformed by an desired unitary operator. For example, if you would like to have a qubit with Hadamard gate applied, rather than teleporting it by a regular EPR pair $(|00\rangle+|11\rangle)$, we teleport it by $(I\otimes H)(|00\rangle+|11\rangle)$. Readers may refer to the original paper of gate teleportation \cite{Gottesman1999} to get a full understanding.

It turns out that cluster state can be used to prepare the resource states for gate teleportation by implementing the one-way scheme. This fact offers a way to define gate fidelity \cite{Fujii2013,Chung2009}. We take Z-rotation gate for example to illustrate this process. The Z-rotation operation is defined as
\begin{equation}
R_{\theta}=\left(\begin{array}{cc}e^{i\frac{\theta}{2}} & 0 \\0 & e^{-i\frac{\theta}{2}}\end{array}\right),
\end{equation}
and the resource state for it is $R_{\theta,2}(|00\rangle +|11\rangle)_{12}$.
\begin{figure}
  \centering
  \includegraphics[width=0.35\textwidth]{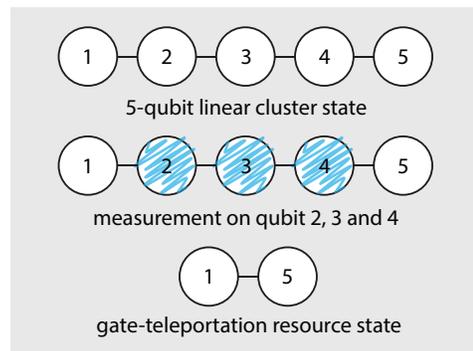}\\
  \caption{(Color online) Using a 5-qubit linear cluster state to produce the gate teleportation resource state for Z-rotation.}\label{gateTele}
\end{figure}

We prepare the resource state from a 5-qubit linear cluster state. The resource state is prepare by a similar measurement sequence to implement a Z-rotation gate in a MBQC system. First, measure qubit 2 on the basis of Pauli $X$ operator. The result state is
\begin{equation}
|\pm\rangle_2 {X_1}^{m_2}(|00\rangle+|11\rangle Z_4)_{13}(|0\rangle+|1\rangle Z_5)_4(|0\rangle+|1\rangle)_5.
\end{equation}
When the out come of the measurement is $|\psi\rangle_2=|+\rangle$, $m_2=0$; when the out come is $|\psi\rangle_2=|-\rangle$, $m_2=1$. Then, we measure qubit 3 on the basis of
\begin{equation}
\cos\tilde{\theta}X+\sin\tilde{\theta}Y=e^{-i \frac{\tilde{\theta}}{2} Z} X e^{i \frac{\tilde{\theta}}{2} Z}.
\end{equation}
Here, $\tilde{\theta}=\pm\theta$, when $|\psi\rangle_2=|\pm\rangle$ respectively. The eigenstate of this operator is
 \begin{eqnarray}
|\varphi_+\rangle &=&\cos\frac{\tilde{\theta}}{2} |+\rangle-i \sin\frac{\tilde{\theta}}{2} |-\rangle\\
|\varphi_-\rangle &=&\cos\frac{\tilde{\theta}}{2} |-\rangle-i \sin\frac{\tilde{\theta}}{2} |+\rangle.
\end{eqnarray}
After the measuring on qubit 3, the system becomes
\begin{eqnarray}
& &|\varphi_\pm\rangle_3 {X_1}^{m_2}{Z_1}^{m_3}(|0\rangle e^{i\frac{\tilde{\theta}}{2}}+|1\rangle e^{-i\frac{\tilde{\theta}}{2}} Z_4)_{1}\nonumber\\
& &(|0\rangle+|1\rangle Z_5)_4(|0\rangle+|1\rangle)_5\nonumber\\
&=&|\varphi_\pm\rangle_3 {X_1}^{m_2}{Z_1}^{m_3}R_{\tilde{\theta},1}(|0\rangle +|1\rangle Z_4)_{1}\nonumber\\
& &(|0\rangle+|1\rangle Z_5)_4(|0\rangle+|1\rangle)_5.
\end{eqnarray}
Next, measure qubit 4 on the $X$ basis. The result state will be
\begin{eqnarray}
& &|\pm\rangle_4 {X_1}^{m_2}{Z_1}^{m_3}R_{\tilde{\theta},1}{X_1}^{m_4}(|00\rangle +|11\rangle)_{15}\nonumber\\
&=&|\pm\rangle_4 ({X_1}^{m_2}R_{\tilde{\theta},1}{X_1}^{m_2}){X_1}^{m_2}{Z_1}^{m_3}{X_1}^{m_4}(|00\rangle +|11\rangle)_{15}\nonumber\\
&=&|\pm\rangle_4 R_{\theta,1}{X_1}^{m_2}{Z_1}^{m_3}{X_1}^{m_4}(|00\rangle +|11\rangle)_{15}.
\end{eqnarray}
We rewrite the state of qubits 1 and 5 as
\begin{eqnarray}
{X_5}^{m_4}{Z_5}^{m_3}{X_5}^{m_2}R_{\theta,5}(|00\rangle +|11\rangle)_{15}.
\end{eqnarray}
After correcting Pauli errors, $B_\mathbf{m}={X_5}^{m_2}{Z_5}^{m_3}{X_5}^{m_4}$ on qubit 5, we get the resource state for Z-rotation.

We define fidelity for a gate operation as
\begin{equation}\label{originalGateFid}
F_U=\Tr\rho_U|\Psi_U\rangle\langle\Psi_U|,
\end{equation}
where $|\Psi_U\rangle$ is the perfect resource state, and $\rho_U$ is the state we measure. From our example we know that
\begin{equation}
\rho_U=\Tr_p\sum_\mathbf{m} B_\mathbf{m}P_\mathbf{m} \rho P_\mathbf{m} {B_\mathbf{m}}^\dagger,
\end{equation}
where $\mathbf{m}$ is the measuring outcome sequence, $P_\mathbf{m}$ is the projection operator, and $B_\mathbf{m}$ is the error-correction operator. $\Tr_p$, the partial trace operation, trace out qubits other than the qubits in the resource state. Back to our Z-rotation resource state. $P_\mathbf{m}$ stands for the measurement sequence on qubit 2, 3, and 4, and $B_\mathbf{m}$ stands for the correction operator ${X_5}^{m_2}{Z_5}^{m_3}{X_5}^{m_4}$. After the measurement and Pauli-error correction, a partial trace is taken so only qubit 1 and 5 exist in $\rho_U$.

If our qubit chain is in a perfect cluster state, we will have $\rho_U=|\Psi_U\rangle\langle\Psi_U|$, so our fidelity defined above results $1$ in this case, coinciding with the perfect application of a Z-rotation operation. As the supplementary material of Ref. \cite{Fujii2013} points out, Jamiolkowski isomorphism ensures the correctness of gate fidelity such defined.

At last, it is enlightening to write the gate fidelity as
\begin{equation}
F_U= {\rm Tr} \left(\rho_U\frac{S_1+I}{2} \frac{S_2+I}{2} \right),\label{fidel}
\end{equation}
where $S_1$ and $S_2$ are stabilizers of $|\Psi_U\rangle$. Equation (\ref{fidel}) can be further simplified for each type of gate (see equation (18)-(21) in the supplementary material of Ref. \cite{Fujii2013}), which simplifies calculation greatly.

\section{CZ-gate Creation Scheme}\label{gatecreation}
In 2001, Briegel and Raussendorf proposed the first scheme to create cluster states \cite{Briegel2001}.
First, prepare all qubits in the state $|+\rangle$. Then, entangle them by applying Ising Hamiltonian for a time interval, of which the accumulated effect is a controlled-Z (CZ) operation for each pair of neighboring sites. After these two steps, a cluster state is prepared. In fact, one can easily generalize this preparation: whatever applies CZ operations to all neighboring sites can fulfill this scheme.

The cluster state prepared in such way would deteriorate with time. We would study the deterioration process in this section.

We assume that the creation process produces a perfect cluster state. After the state is prepared, the system evolves over time and deteriorates because of coupling to a boson environment.

\subsection{Pure Phase Noise: The Exactly Solvable Case}
We first limit the coupling term to pure phase noise. Without the presence of amplitude noise, we can solve the time-evolution problem exactly. We take amplitude noise into consideration in Sec. \ref{partC}. The Hamiltonian here reads

\begin{equation}\label{hamil1}
 H=\sum_{n=1}^{N} \epsilon_n \sigma_z^{(n)} + \sum_\mathbf{k} \omega_\mathbf{k} {a_\mathbf{k}}^\dagger a_\mathbf{k} + \sum_{n,\mathbf{k}} \sigma_z^{(n)} (g_\mathbf{k} a_\mathbf{k}^\dagger + g_\mathbf{k}^\ast a_\mathbf{k}),
 \end{equation}
where $N$ is the qubit number, $\epsilon_n$ is half of the energy gap between $|0\rangle$ and $|1\rangle$ state of the $n$th qubit, and $\sigma_z^{(n)}$ is the Pauli Z operator of the $n$th qubit. $\omega_\mathbf{k}$ is the frequency or energy of a boson mode. We set $\hbar=1$ in this paper. $a_\mathbf{k}^{(\dagger)}$ is the annihilation (creation) operator. The third term is the coupling term, with $g_\mathbf{k}$ being the coupling coefficient. This Hamiltonian is similar to the Dicke Model \cite{Dicke1954}, and the coupling term has been comprehensively studied in single qubit case \cite{Leggett1987}.  Presumably, the coupling is small in quantum computation situations, but our analysis applies to all coupling strengths.

Our qubits are initially prepared in a perfect cluster state:
\begin{equation}
\rho^Q(t=0)=|\Psi_C\rangle\langle\Psi_C|.
\end{equation}
Our boson environment is set in a thermal state in the beginning:
\begin{eqnarray}\label{rhob}
\rho^B(t=0) &=& \frac{\exp\left(-\beta \sum_\mathbf{k} \epsilon_k a_\mathbf{k}^\dagger a_\mathbf{k}\right)}{\Tr\left[\exp\left(-\beta \sum_\mathbf{k} \epsilon_k a_\mathbf{k}^\dagger a_\mathbf{k}\right)\right]}\nonumber\\
&=& \prod_\mathbf{k} \frac{\exp(-\beta \epsilon_k {a_\mathbf{k}}^\dagger a_\mathbf{k})}{1+\langle N_{\omega_k} \rangle},
\end{eqnarray}
where $\beta=1/k_B T$, and $\langle N_{\omega_k}\rangle$ is the mean boson number with frequency $\omega_k$ in thermal state.

We solve the time-evolution problem in the interaction picture.  Choosing $H_0$ as
\begin{equation}
H_0=\sum_{n=1}^{N} \epsilon_n \sigma_z^{(n)} + \sum_\mathbf{k} \epsilon_\mathbf{k} a_\mathbf{k}^\dagger a_\mathbf{k},
\end{equation}
we get the interaction part of the Hamiltonian
\begin{eqnarray}\label{vi}
V_I(t) &=& e^{i H_0 t} \bigg( \sum_{n,\mathbf{k}} \sigma_z^{(n)} (g_\mathbf{k} a_\mathbf{k}^\dagger + g_\mathbf{k}^\ast a_\mathbf{k})\bigg) e^{-i H_0 t} \nonumber\\
&=& \sum_{n,\mathbf{k}} \sigma_z^{(n)} (g_\mathbf{k} e^{i \omega_k t} a_\mathbf{k}^\dagger + g_\mathbf{k}^\ast e^{-i \omega_k t} a_\mathbf{k}).
\end{eqnarray}
The unitary time evolution operator is
\begin{eqnarray}\label{ui}
U_I(t) &=& \hat{T} \exp\left[-i \int_0^t V_I (t^\prime) dt^\prime\right]\nonumber\\
&=& \exp\left\{\sum_{n,\mathbf{k}} \left[g_\mathbf{k} \sigma_z^{(n)} \varphi_{\omega_k}(t) a_\mathbf{k}^\dagger - g_\mathbf{k}^\ast \sigma_z^{(n)} \varphi_{\omega_k}^\ast (t) a_\mathbf{k} \right] \right\}\nonumber\\
& & \times \exp\left\{i \sum_\mathbf{k} \sum_{m,n} |g_\mathbf{k}|^2 \sigma_z^{(m)} \sigma_z^{(n)} s(\omega_k,t)\right\},
\end{eqnarray}
where
\begin{eqnarray}
\varphi_{\omega_k}(t)&=&\frac{1-e^{i \omega_k t}}{\omega_k},\\
s(\omega_k,t)&=&\frac{\omega_k t - \sin(\omega_k t)}{{\omega_k}^2}.
\end{eqnarray}
We present the detailed derivation of this part in Appendix \ref{appAi}. The reduced density matrix of the qubits part evolves like
\begin{equation}\label{qubitEvolve}
\rho^Q_I(t)={\rm Tr_E}\left[U_I(t)\rho^Q_I(0) \otimes \rho^E_I(0) {U_I}^\dagger (t) \right].
\end{equation}
After some calculation, we get the component form of the density operator
\begin{eqnarray}\label{interRho}
\rho^Q_{I,\{i_n,j_n\}}(t)&\equiv&\langle i_1,i_2,\cdots, i_N|\rho^Q_I(t)|j_1,j_2,\cdots,j_n\rangle\nonumber\\
&=& \exp\left\{-\Gamma(t,T)\left[\sum_{n=1}^N(i_n-j_n)\right]^2\right\}\nonumber\\
& &\times \exp\left\{i \Theta(t)\left[ \left(\sum_{n=1}^N i_n\right)^2-\left(\sum_{n=1}^N j_n\right)^2 \right] \right\}\nonumber\\
& & \times \rho_{I,\{i_n,j_n\}}^Q(0).\label{rhoi}
\end{eqnarray}
The subscript $I$ means interaction picture. The expression of $\Gamma(t,T)$ and $\Theta(t)$, and the detailed calculation of equation (\ref{qubitEvolve}) is presented in Appendix \ref{appAii}. We proceed to take the continuum limit in the appendix, after which $g_\mathbf{k}$ contained in the spectral density. Afterwards, we assume an ohmic spectral density here
\begin{equation}
I(\omega)=\eta \omega e^{-\omega/\omega_c}.
\end{equation}
After some calculation, we reach at
\begin{equation}\label{ohmic1}
\Gamma(t,T)=\eta \ln(1+ {\omega_c}^2 t^2)+\eta \ln\left(\frac{\beta}{\pi t} \sinh \frac{\pi t}{\beta}\right),
\end{equation}
\begin{equation}\label{ohmic2}
\Theta(t)=\eta \omega_c t- \eta \arctan(\omega_c t).
\end{equation}

\begin{figure}
  \centering
  \subfigure[]{
    \includegraphics[width=4.1cm]{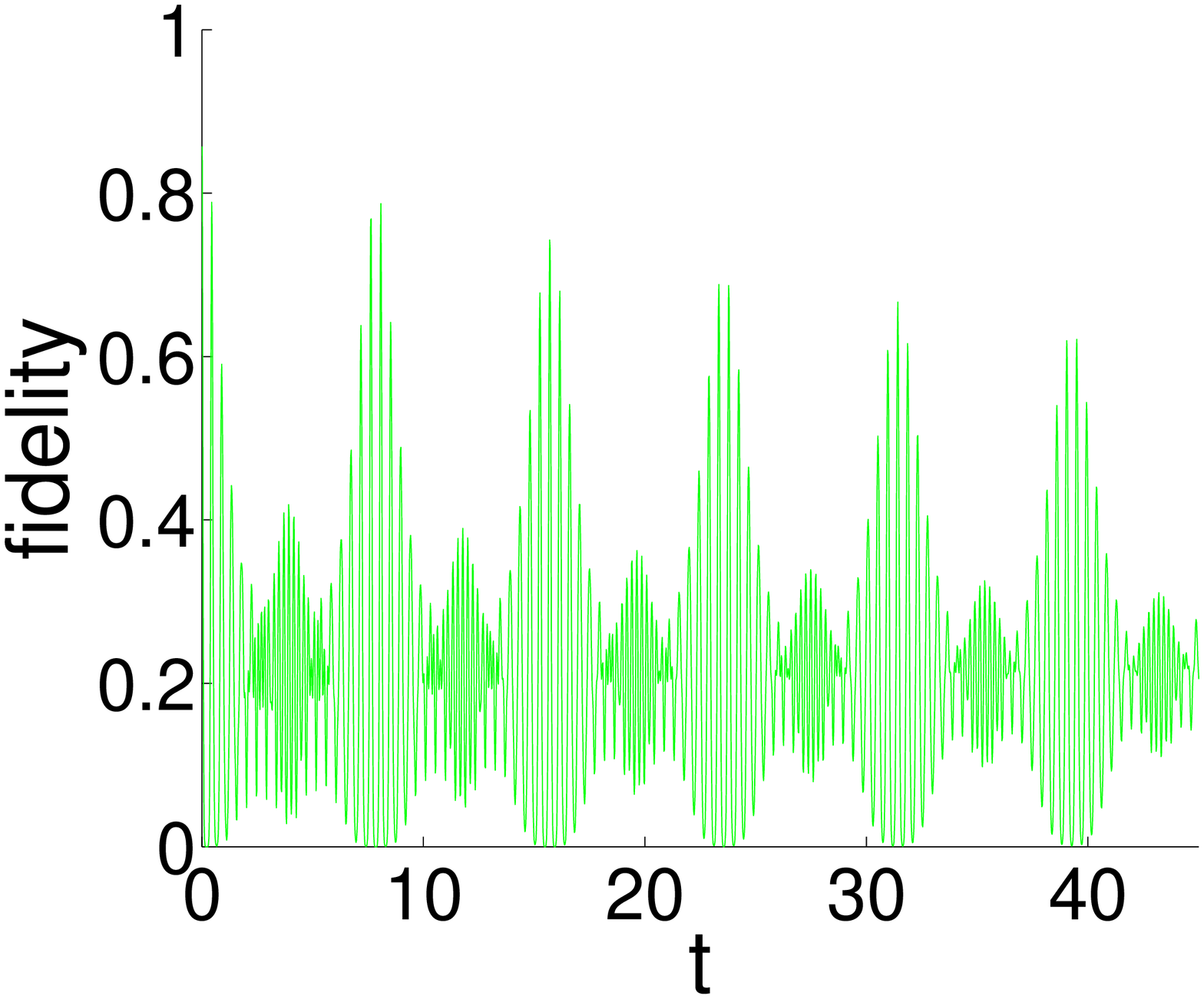}}
    \subfigure[]{
    \includegraphics[width=4.1cm]{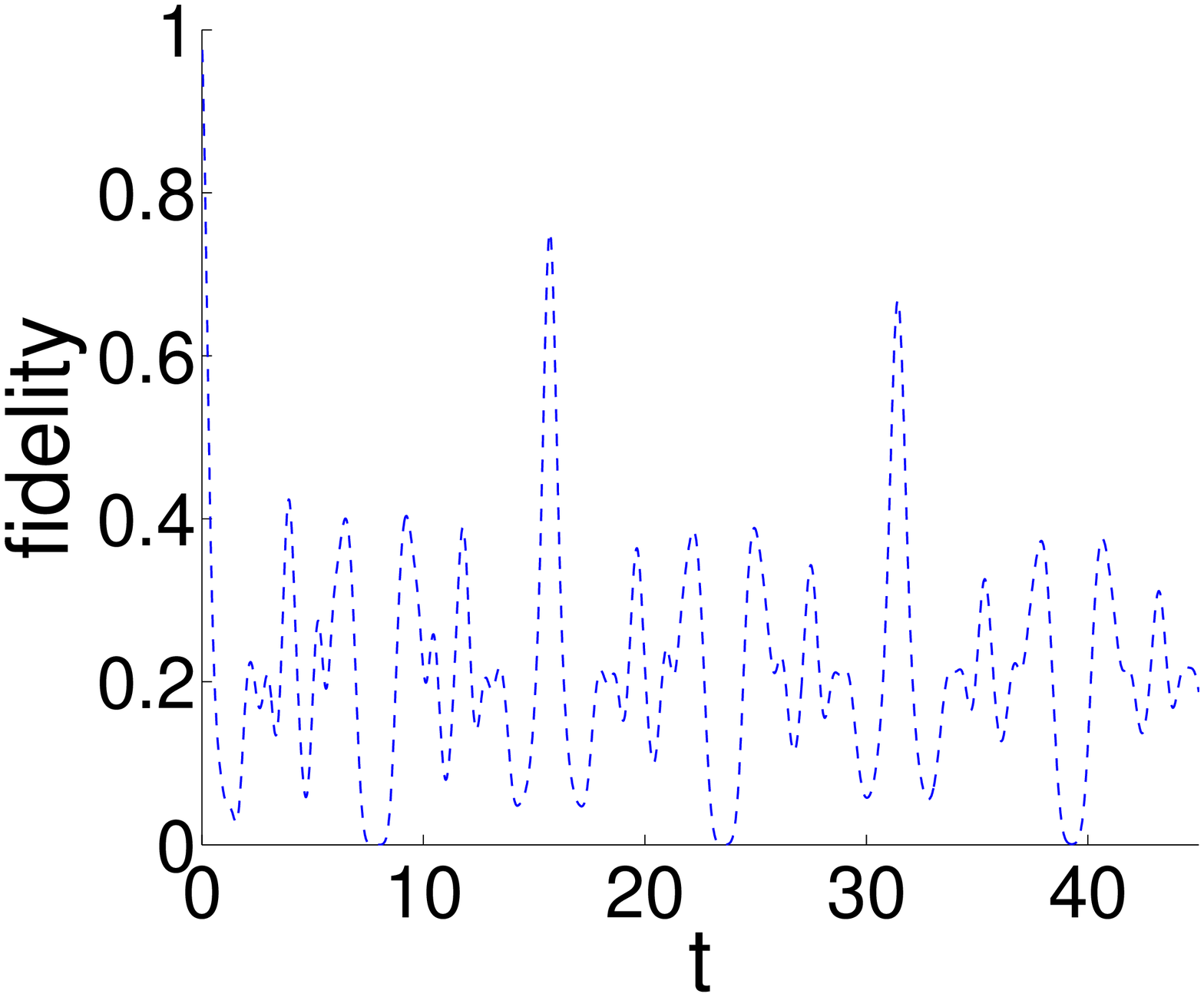}}
  \subfigure[]{
    \includegraphics[width=4.1cm]{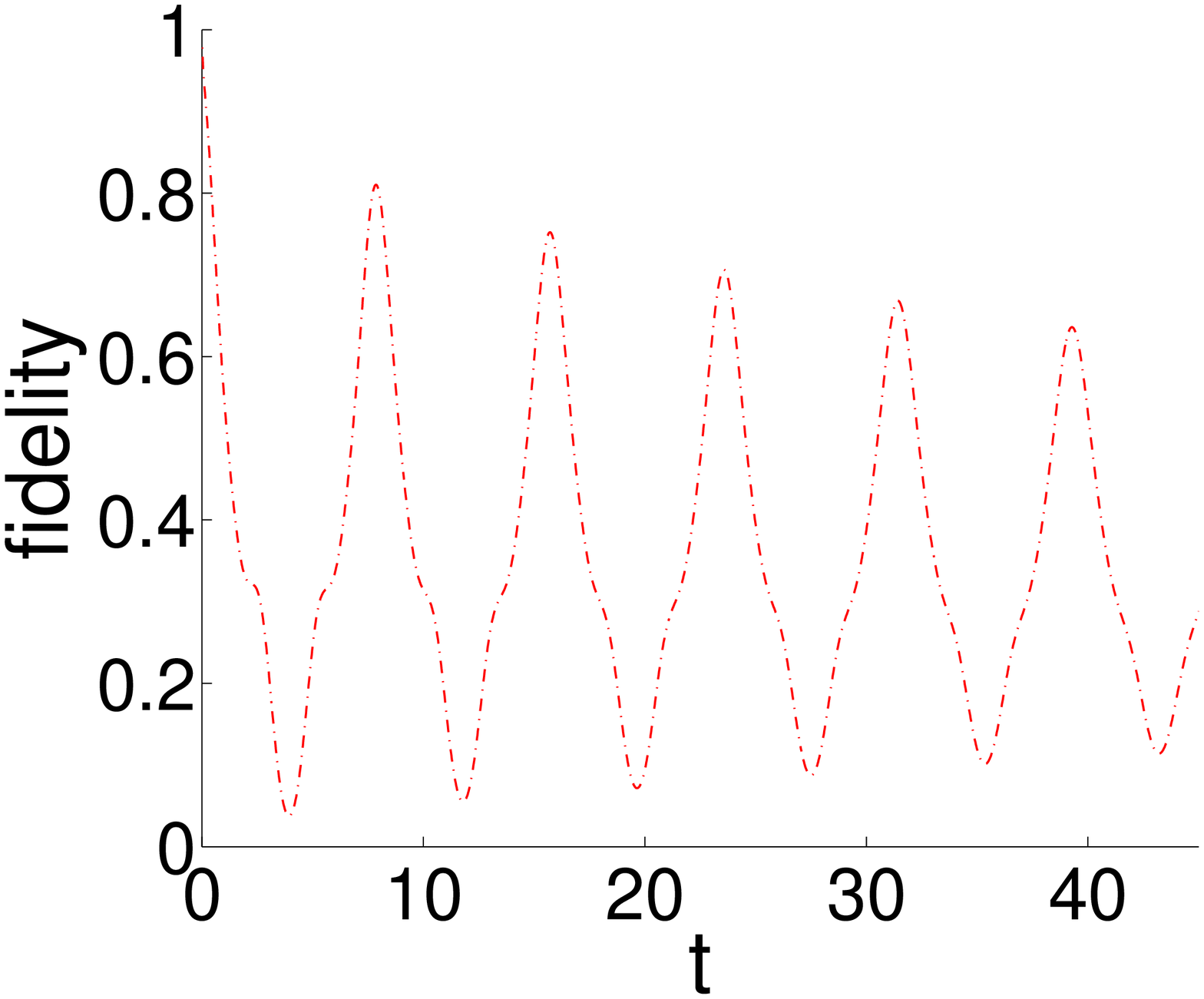}}
  \subfigure[]{
    \includegraphics[width=4.1cm]{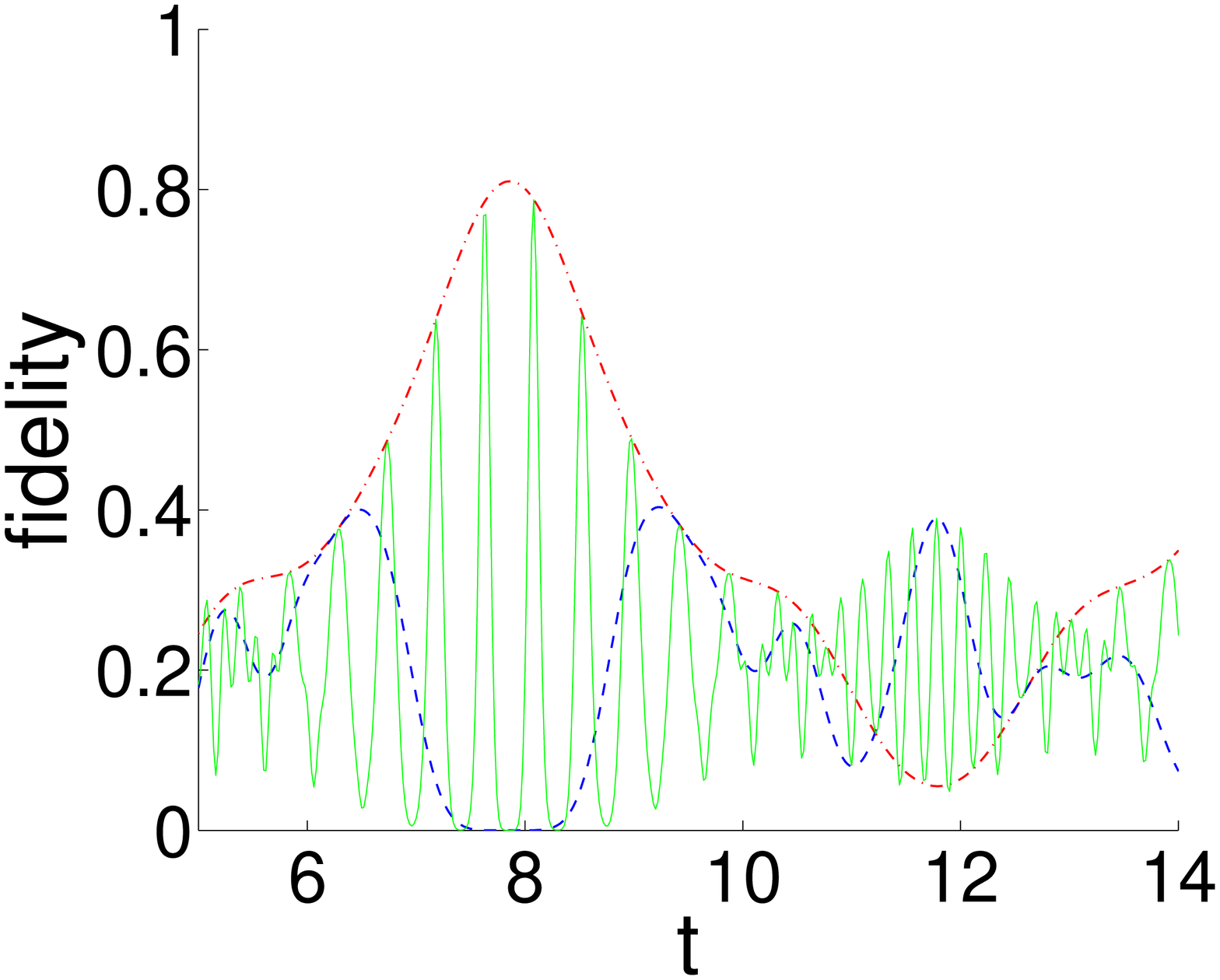}}
  \caption{(Color online) Fidelity-time relation of a 7-qubit linear cluster state. Solid green lines for $\epsilon_n=3$, dashed blue lines for $\epsilon_n=0.9$, and dash-dotted red lines for $\epsilon_n=0$. (a) $\epsilon_n=3$. (b) $\epsilon_n=0.9$. (c) $\epsilon_n=0$. (d) Zooming in to the whole figure, with three $\epsilon_n$ all presented.}
  \label{cluster}
\end{figure}
The fidelity of a cluster state here is defined as
\begin{equation}\label{fidelityy}
F = {\rm Tr} \left(|\Psi_C\rangle\langle\Psi_C|\rho^Q(t)\right).
\end{equation}
Here $\rho^Q$ is the reduced density operator of the qubits. Remember that if the perfect cluster state is in Schr\"odinger picture, we of course require $\rho_Q$ presented in Schr\"odinger picture. Rewriting the equation using the density operator in interaction picture, we have
\begin{eqnarray}\label{fidInteraction}
F(t) =  {\rm Tr} \left(|\Psi_C\rangle\langle\Psi_C|e^{-i H_0 t} \rho_I^Q(t) e^{i H_0 t}\right).\label{right}
\end{eqnarray}

We plot a 7-qubit linear cluster state coupled with boson environment (Fig. \ref{cluster}). The parameters are set to $\eta=1/1000$, $\omega_c=100$, and $\beta/\pi=1$. There are three lines in the figure, green for $\epsilon_n=3$, blue for $\epsilon_n=0.9$, and red for $\epsilon_n=0$.

There are two factors contributing to the oscillation of fidelity over time: the $\Theta$ function and the $H_0$ part. If the two oscillation frequencies are close, the oscillation pattern is highly unpredictable (see the dashed blue lines). To any real systems, this case should be avoided. We emphasize that when the coupling does not exist, the fidelity still oscillates due to the $H_0$ part, but the peak of the oscillation is always $1$. Another fact is that even when the temperature of boson environment is zero, there still exists a fidelity drop at the peak. This is easy to understand, since that the coupling term still works to the system, even though all modes of boson environment is in the vacuum state.

\begin{figure}
  \centering
  \subfigure[]{
    \includegraphics[width=4.1cm]{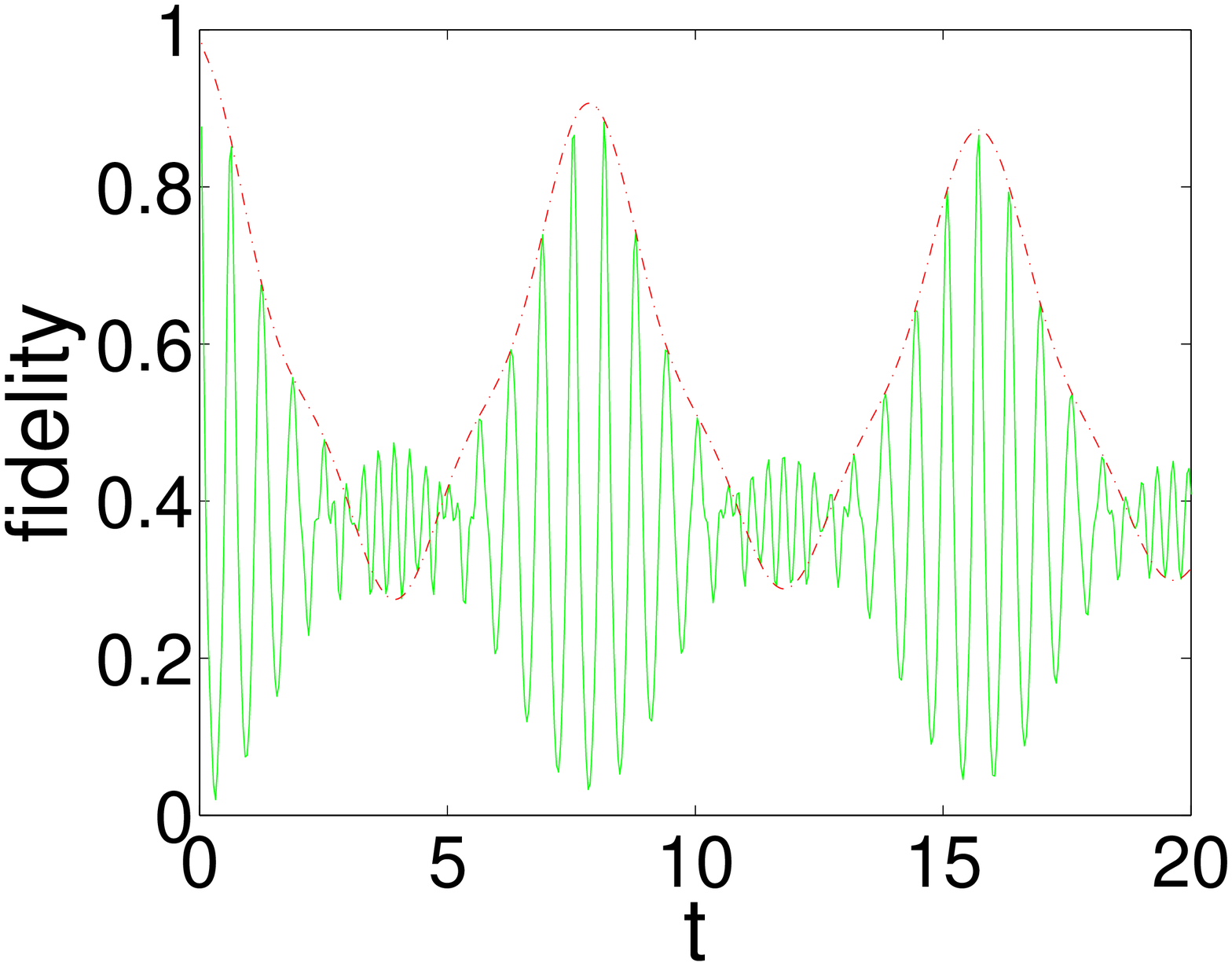}}
  \subfigure[]{
    \includegraphics[width=4.1cm]{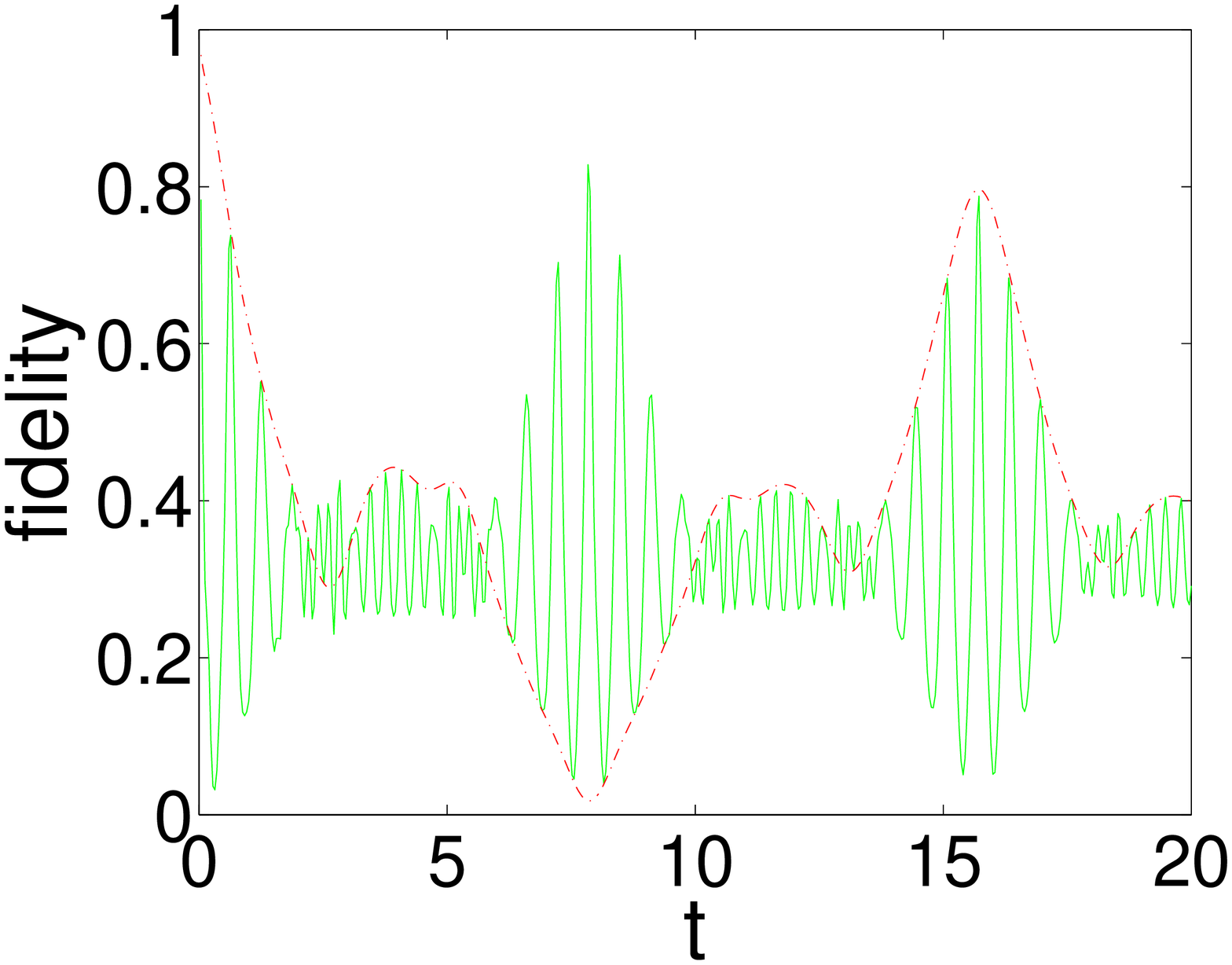}}\\
  \subfigure[]{
    \includegraphics[width=4.1cm]{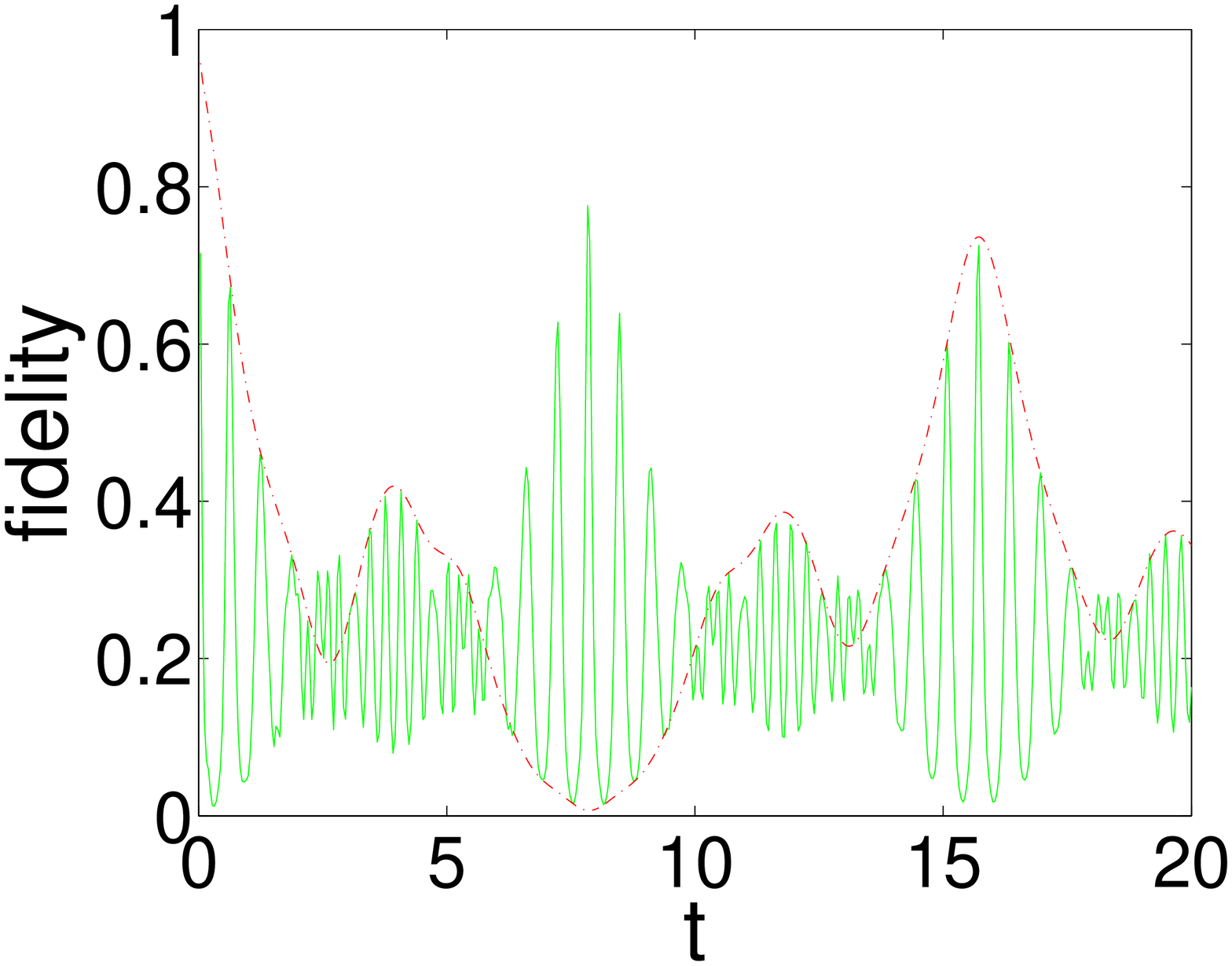}}
  \subfigure[]{
    \includegraphics[width=4.1cm]{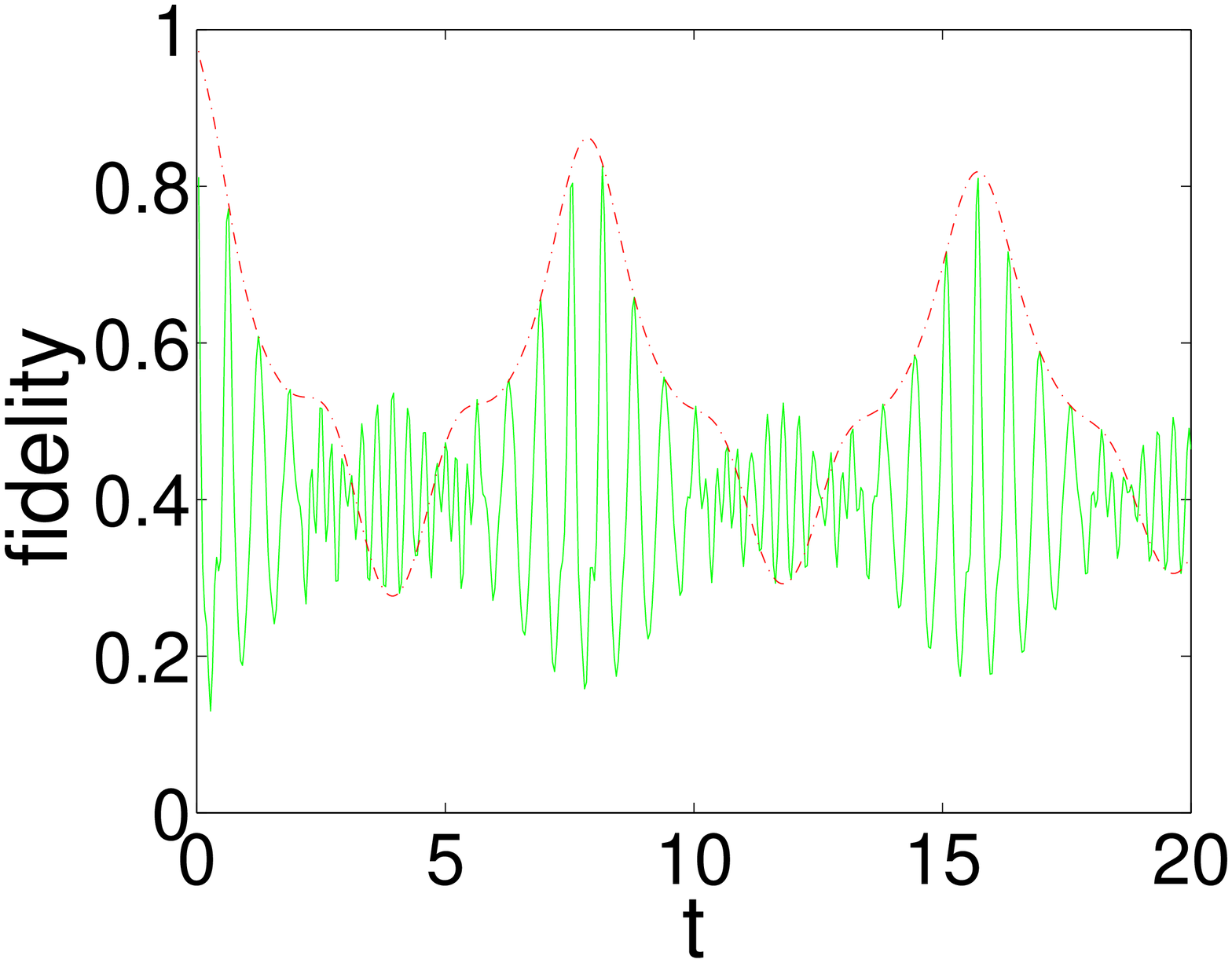}}
  \caption{(Color online) Fidelity-time dependence. Solid green lines for $\epsilon_n=5$; dash-dotted red lines for $\epsilon_n=0$. (a) 5-qubit identity gate. (b) 8-qubit Hadamard gate. (c) CZ gate. (d) Z-rotation gate, $\zeta=\pi/8$.}
  \label{way1}
\end{figure}

To be more specific, we analyze fidelity for various gate operations:
\begin{eqnarray}
F_U= {\rm Tr} \left(\rho_U(t) |\Psi_U\rangle \langle \Psi_U| \right).
\end{eqnarray}
The fidelity-time dependence for 5-qubit identity gate, 8-qubit Hadamard gate, Z-rotation gate and Controlled-Z gate are plotted in Fig. \ref{way1}.
The parameters are $\eta=1/1000$, $\omega_c=100$, and $\beta/\pi=1$. Because a medium $\epsilon_n$ will result in unpredictable time evolution pattern, this kind of fidelity curve is not presented. Different gate operations show different oscillation patterns, due to their own lattice structures and sizes.

Fig. \ref{way1} shows that when $\epsilon_n$ is large, all types of gates exhibit a fidelity peak uniformly at $t \approx 8$, which is better than $\epsilon_n=0$ case. This is because when $\epsilon_n$ is large, the $H_0$ renders the fidelity to oscillate rapidly, which leave the gate-type-independent $\Theta$ function to shape the envelope of the fidelity. Therefore, if the measurement can be implemented rapidly, a strong $\epsilon_n$ is preferred.

\subsection{Suggestions to Enhance the Performance of CZ-Gate Creation Scheme}
Because of the advantage listed above, we pick the green line of Fig. \ref{way1} to conduct our investigation in this part. In this part, we are focused on giving suggestions to enhance the performance of MBQC realized by the CZ-creation scheme.

\begin{figure}
  \centering
  \subfigure[]{
    \label{deriRhos} 
    \includegraphics[width=4.1cm]{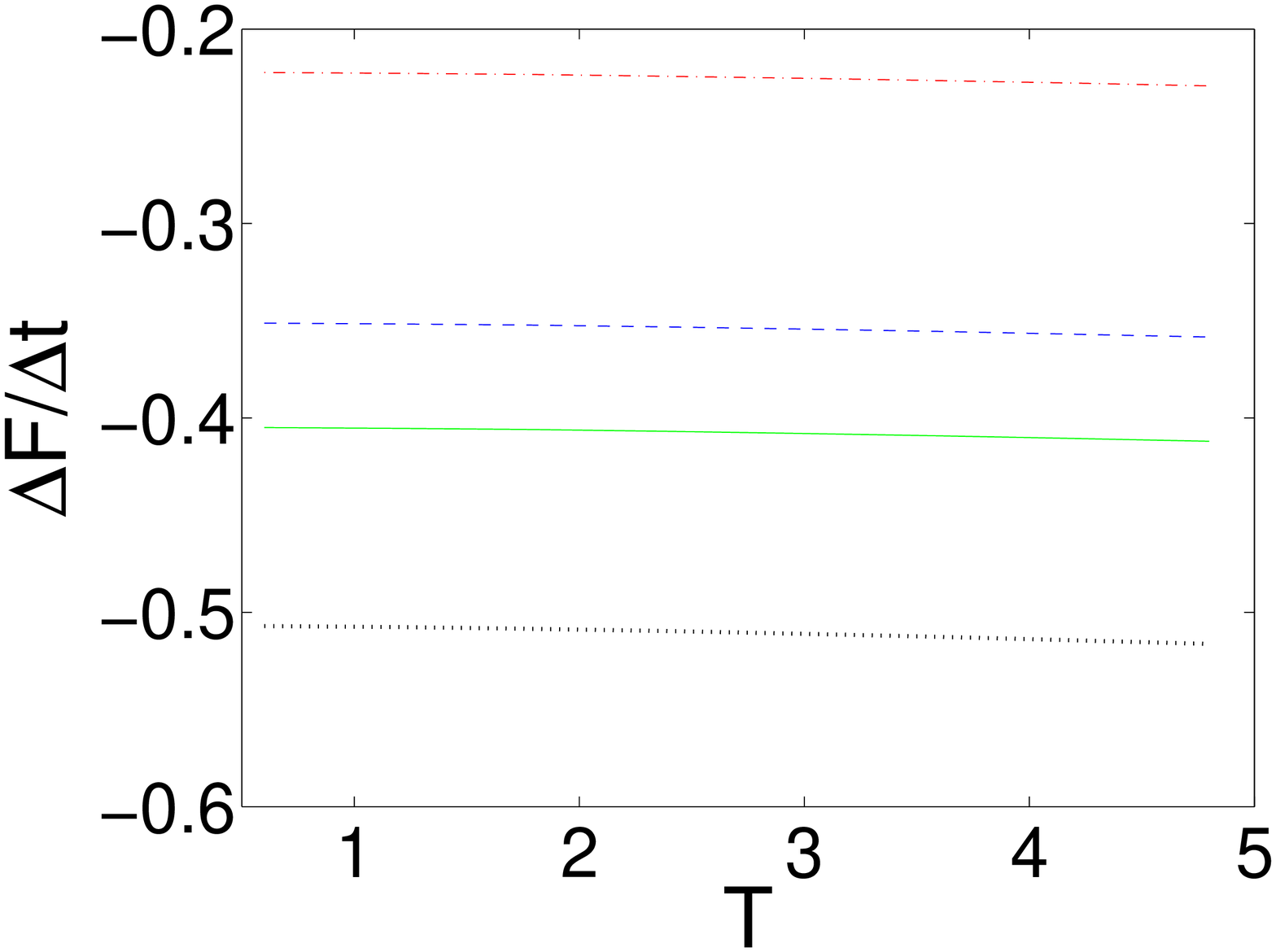}}
  \subfigure[]{
    \label{deriId5} 
    \includegraphics[width=4.1cm]{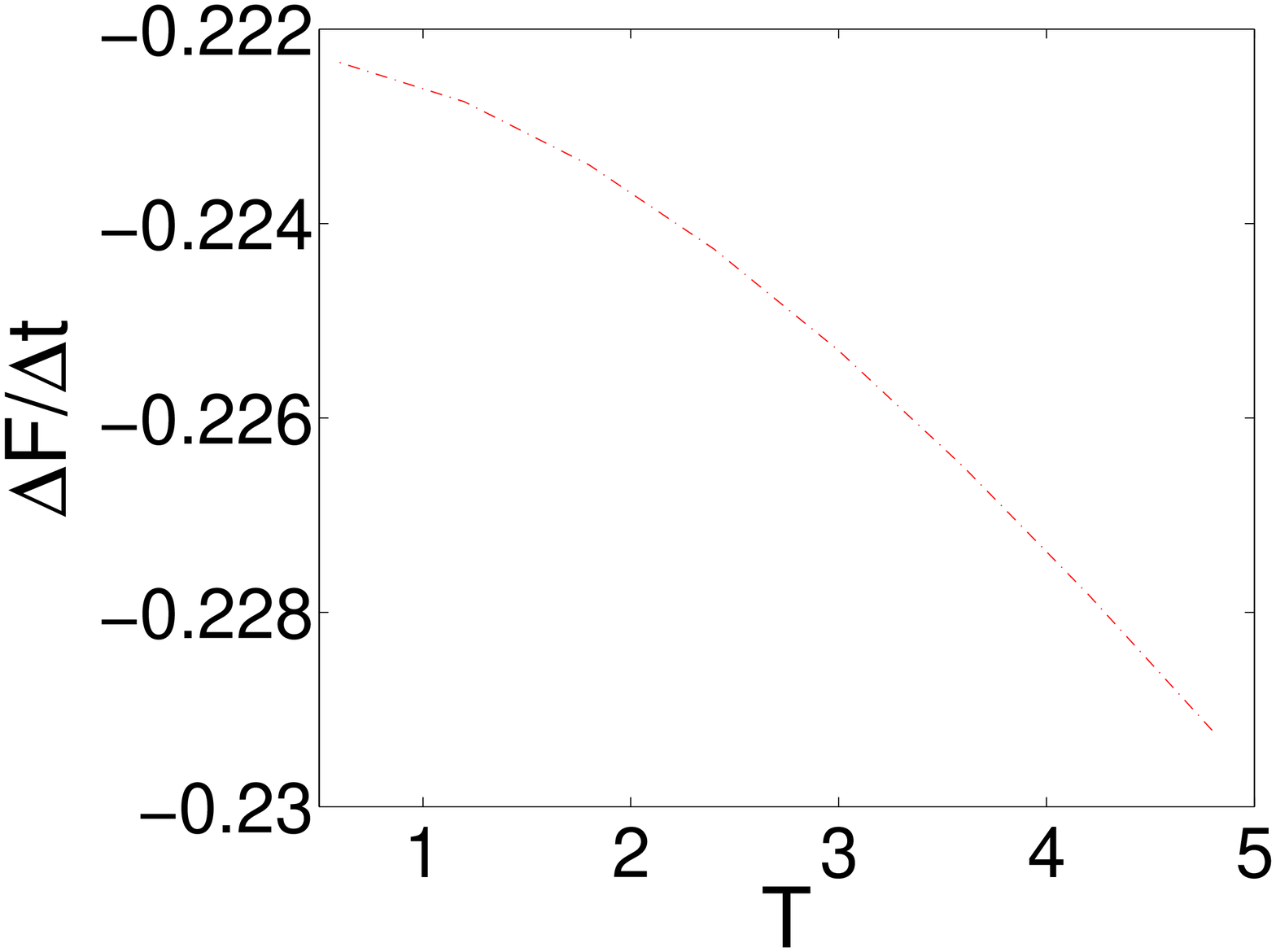}}
  \caption{(Color online) The derivative of fidelity as the function of temperature. Dash-dotted red lines for 5-qubit identity gate, solid green lines for 8-qubit Hadamard gate, dashed blue lines for Z-rotation gate and dotted black lines for CZ gate. (a) The derivative of four gate operations. (b) Extraction: the derivative of 5-qubit identity gate only.}
  \label{derivative}
\end{figure}

If the gate operation is implemented as soon as the cluster state is prepared, one would concern the drop rate of the fidelity on the first several fidelity peaks. We treat the fidelity drop rate as
\begin{equation}
\frac{\Delta F}{\Delta t}\bigg|_{t \rightarrow 0} = \frac{F(t=t_1)-F(t=0)}{t_1},
\end{equation}
where $t_1$ is the time for the first peak of fidelity after $t=0$. We calculate the fidelity drop rate versus the temperature of the boson environment. Our result is shown in Fig. \ref{derivative}.
As the figure reveals, the fidelity drop rate curves decrease as the temperature increases, but the decrease is slow and non-linear. When the environment temperature gets lower, the decrease becomes even slower. This process eventually stops at $T=0$, where the fidelity drop rates reach at a non-zero value. Utilizing this fact, we can save cooling equipments, since there is little room for the fidelity drop rates to decrease in low temperature.

If a cluster state will be kept for some time before measuring it, we would ask where the fidelity peaks become high enough. This question fits the situation when a part of computation must wait until other parts finish, while the cluster state is already prepared. From Fig. \ref{way1} we learn that the fidelity peaks in the first peak of the envelope (see the set of fidelity peaks near $t \approx 8$) may be qualified for our purpose. To evaluate this area, we calculate the arriving time of the highest peak in it, and the fidelity of the highest peak.
\begin{figure}
  \centering
  \subfigure[]{
    \label{ttdepend} 
    \includegraphics[width=4.1cm]{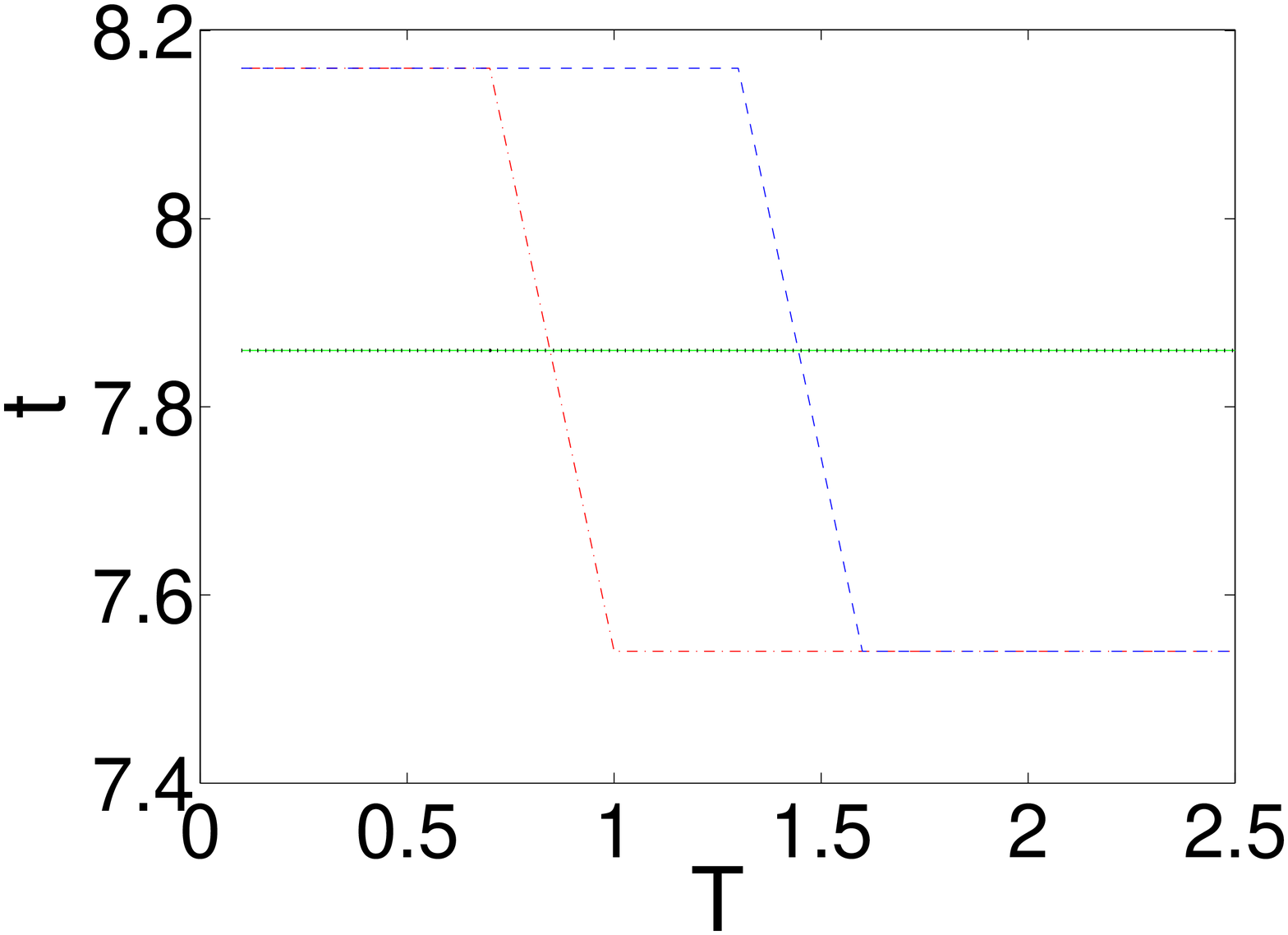}}
  \subfigure[]{
    \label{ftdepend} 
    \includegraphics[width=4.1cm]{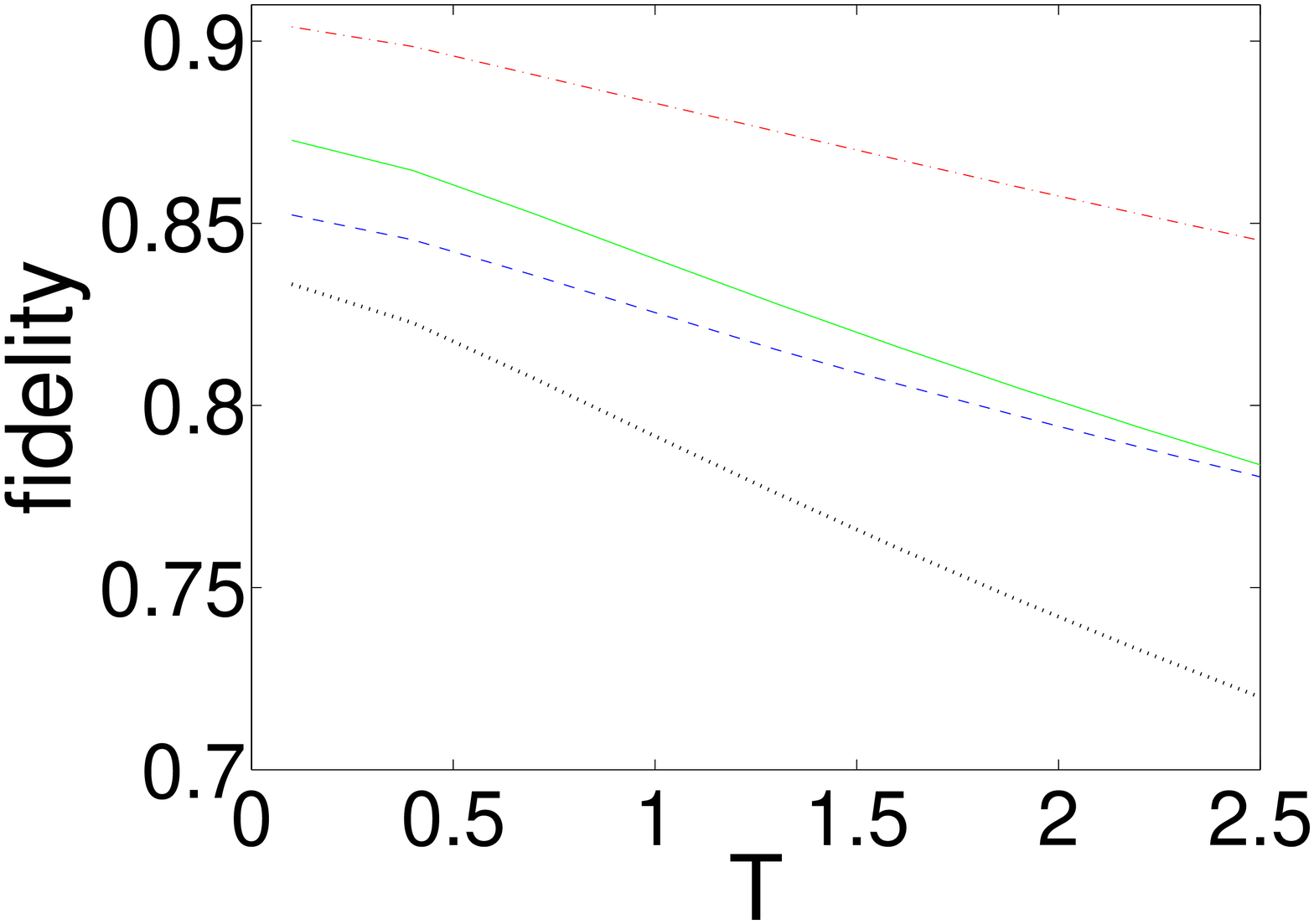}}
  \caption{(Color online) Peak statistics. Dash-dotted red lines for 5-qubit identity gate, solid green lines for 8-qubit Hadamard gate, dashed blue lines for Z-rotation gate and dotted black lines for CZ gate. (a) The arriving time of peak depending on temperature, with the solid green line and the dotted black line overlapped. (b) The peak fidelity depending on temperature.}
  \label{tdepend}
\end{figure}

Our calculation (see Fig. \ref{tdepend}) proves the qualification of this area in two aspects.

\begin{figure}
  \centering
  \subfigure[]{
    \includegraphics[width=4.1cm]{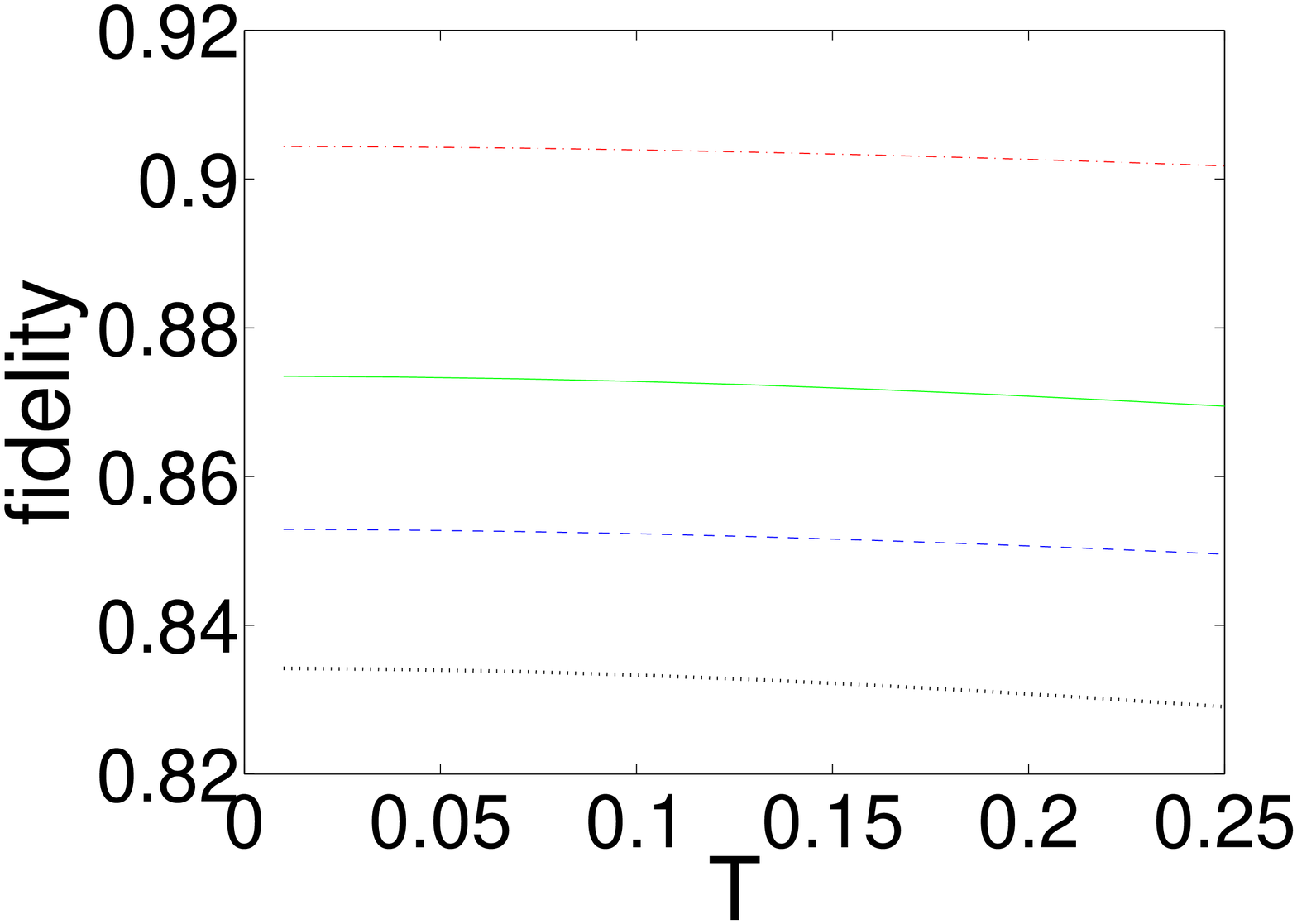}}
  \subfigure[]{
    \includegraphics[width=4.1cm]{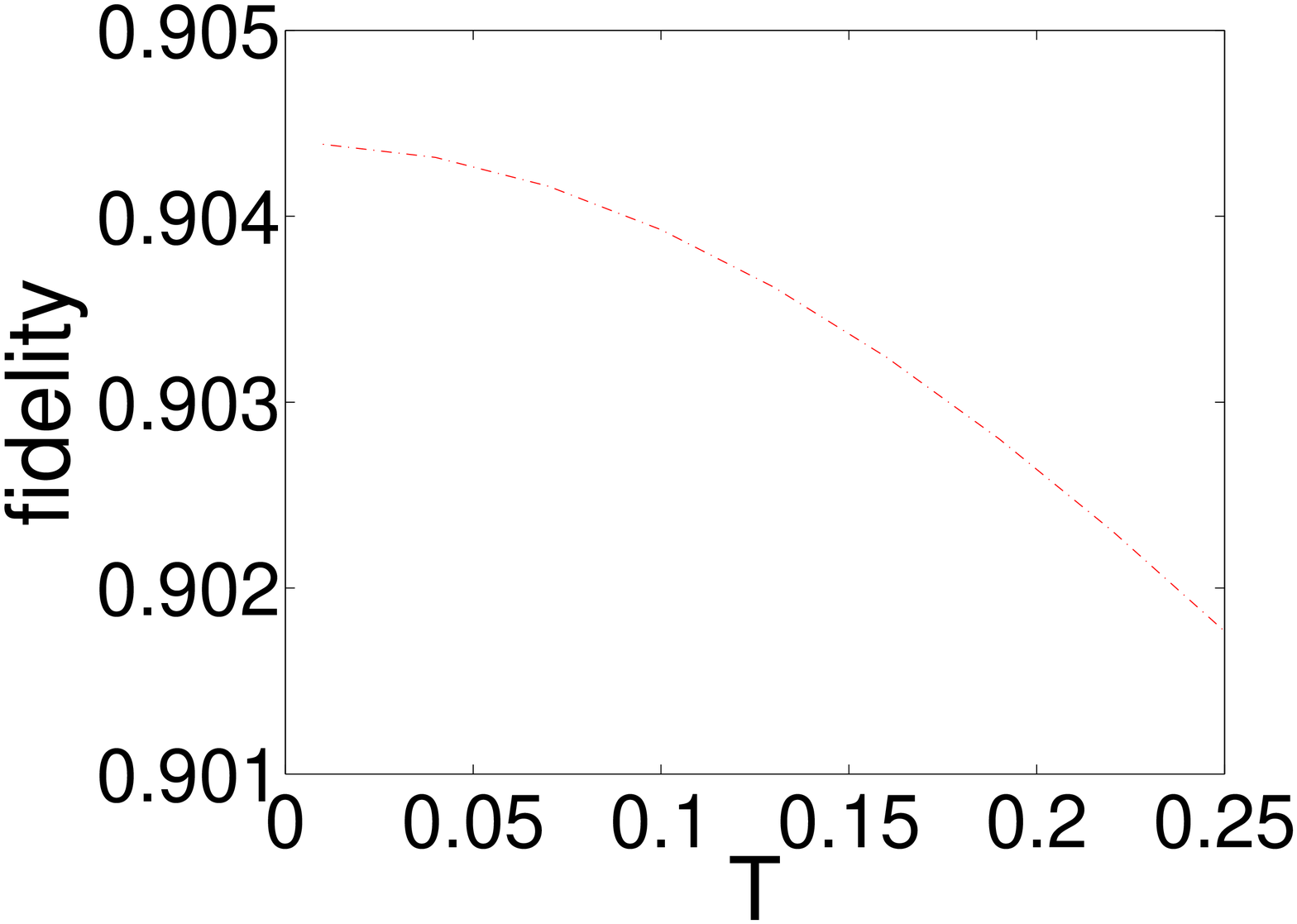}}
  \caption{(Color online) Fidelity-temperature dependence. Dash-dotted red lines for 5-qubit identity gate, solid green lines for 8-qubit Hadamard gate, dashed blue lines for Z-rotation gate and dotted black lines for CZ gate. (a) Four types of gates. (b) Extraction: 5-qubit identity gate only.}
  \label{fidelityPrecise}
\end{figure}

First, the fidelity of the highest peak is high enough. From Fig. \ref{tdepend}, we see that the fidelities are all above $0.8$ when $T<0.5$. We also work out the fidelity-temperature dependence at very low temperature, see Fig. \ref{fidelityPrecise}. When temperature gets even lower, the fidelity changes no more, stoping at a value lower than $1$ at $T=0$. Therefore, we again need not cool down the temperature with infinite effort. In our parameter setting, $\omega_T=0.1$ is good enough to ensure a high fidelity for the highest peak.

Second, the arriving time of the highest peak is almost independent of environment temperature, and completely gate-typing independent, which simplifies our utilization. The highest peak may switch from one to another among several neighboring peaks, as happens on the identity gate and the Z-rotation gate in the figure, but this fact does not complicate our utilization. Since several peaks near the highest peak are almost equally good and the temperature should vary little, we can just stick on one of the peaks. The fact that the highest peak arrives uniformly regardless of $T$ is because the oscillation is mainly controlled by $\Theta$ function and $H_0$, both of which do not depend on the environment temperature at all. The arriving time is also gate-type independent. Again, it is because of the $\Theta$ function, which keeps invariant under changes of gate type, qubit number, and the lattice shape of a cluster state.

\subsection{Generalized Noise: Numerical Results}\label{partC}
Until now, we only evaluated pure phase noise. In the perspective of noise theory, pure phase noise fails to describe all circumstances. As a result, we generalize our Hamiltonian to consider both phase and amplitude noise in this section:
\begin{eqnarray}\label{exampleHamil}
& &H=\sum_{n=1}^{N} \epsilon_n \sigma_z^{(n)} + \sum_\mathbf{k} \omega_k a_\mathbf{k}^\dagger a_\mathbf{k}\nonumber\\
&&+ \sum_{n,\mathbf{k}} (\cos(\theta)\sigma_z^{(n)}-\sin(\theta)\sigma_x^{(n)}) (g_\mathbf{k} a_\mathbf{k}^\dagger + g_\mathbf{k}^\ast a_\mathbf{k}).
\end{eqnarray}
With amplitude noise added, we can no longer solve the time evolution problem analytically. Instead, we seek for numerical solutions, trying to figure out the character of the generalized noise. For simplicity, we calculate a single-mode boson environment, with the frequency resonant with the energy gap of a two-level qubit:
\begin{eqnarray}
& &H=\epsilon \sum_{n=1}^{N} \sigma_z^{(n)} + 2\epsilon a^\dagger a\nonumber\\
&&+ g ( a^\dagger + a)\sum_{n} (\cos(\theta)\sigma_z^{(n)}-\sin(\theta)\sigma_x^{(n)}).
\end{eqnarray}
This single-frequency model is reasonable, because naturally the resonant frequency mode causes more damage. The only drawback of this model is that it fails to describe the situation when $\epsilon=0$, in which case a boson would have zero energy, which is impossible.

For cluster states consisting of several qubits, we write a program to calculate the time evolution of the density operator:
\begin{equation}
\rho^Q(t)=e^{-iHt}\rho^Q(0) e^{iHt}
\end{equation}

In the system, the boson environment is infinite dimensional. In our program, however, we adopt the cut-off approximation, setting the maximum boson number to some large number.

\begin{figure}
  \centering
  \subfigure[]{
    \includegraphics[width=4.1cm]{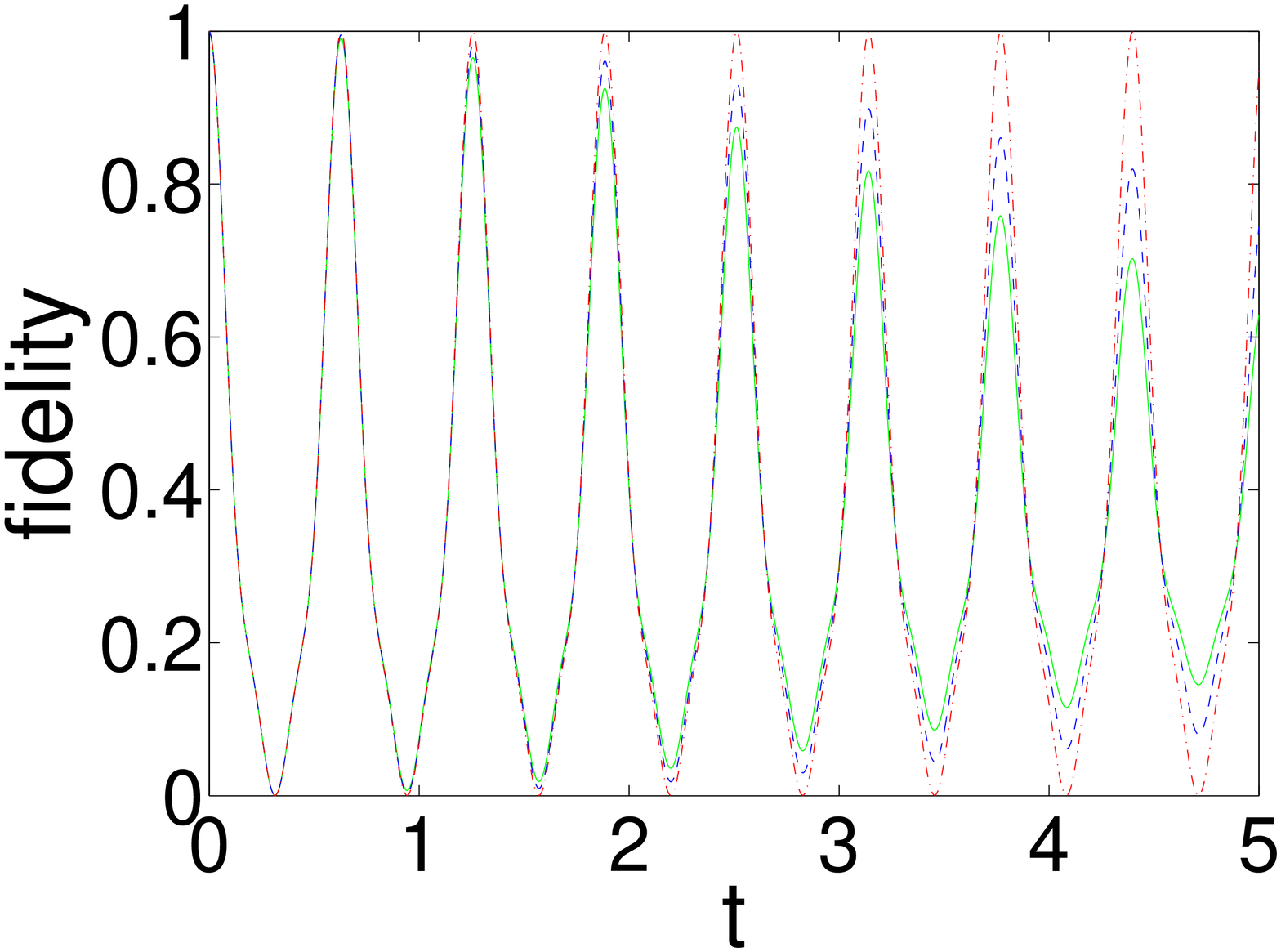}}
  \subfigure[]{
    \includegraphics[width=4.1cm]{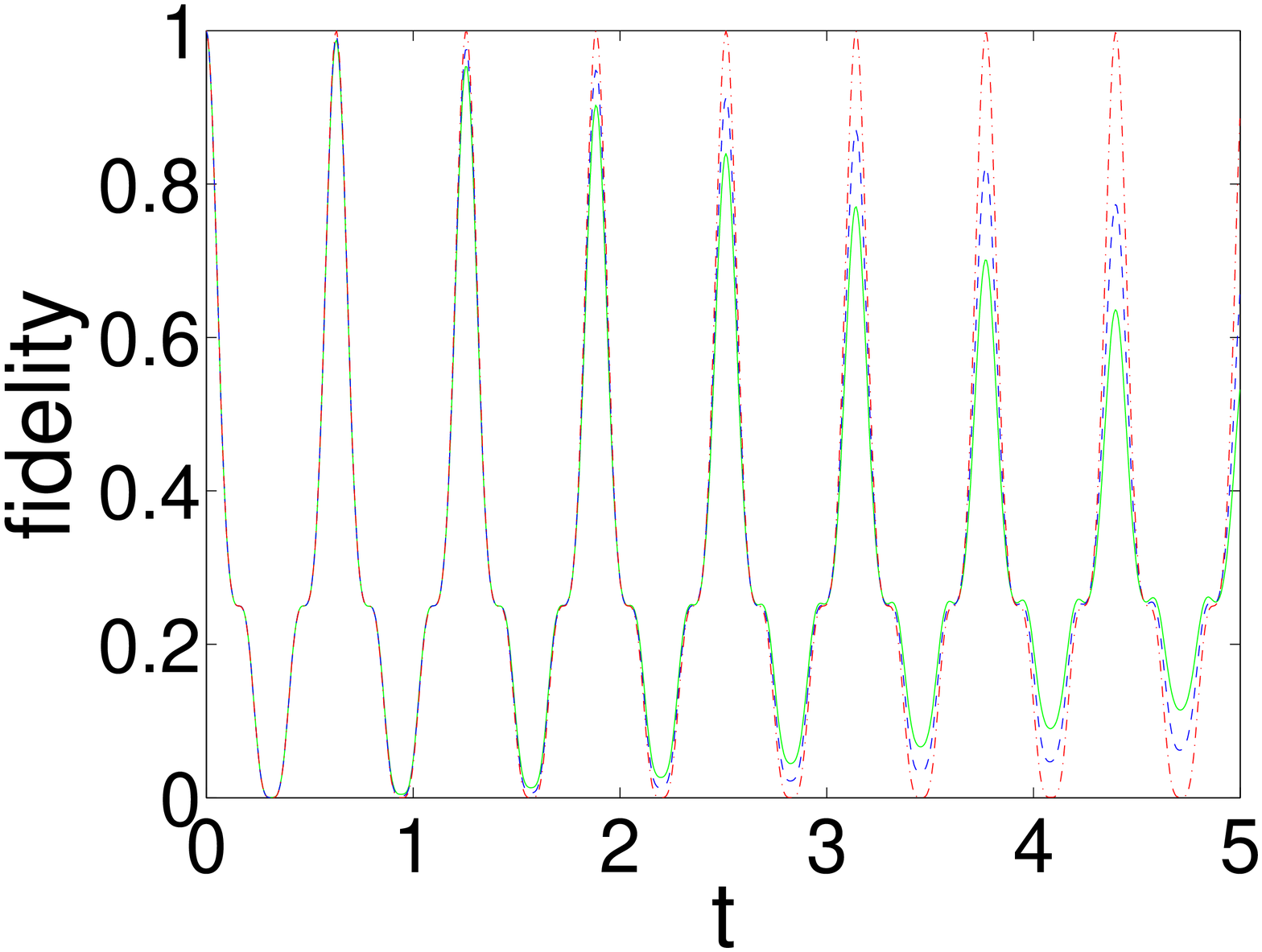}}\\
  \subfigure[]{
    \includegraphics[width=4.1cm]{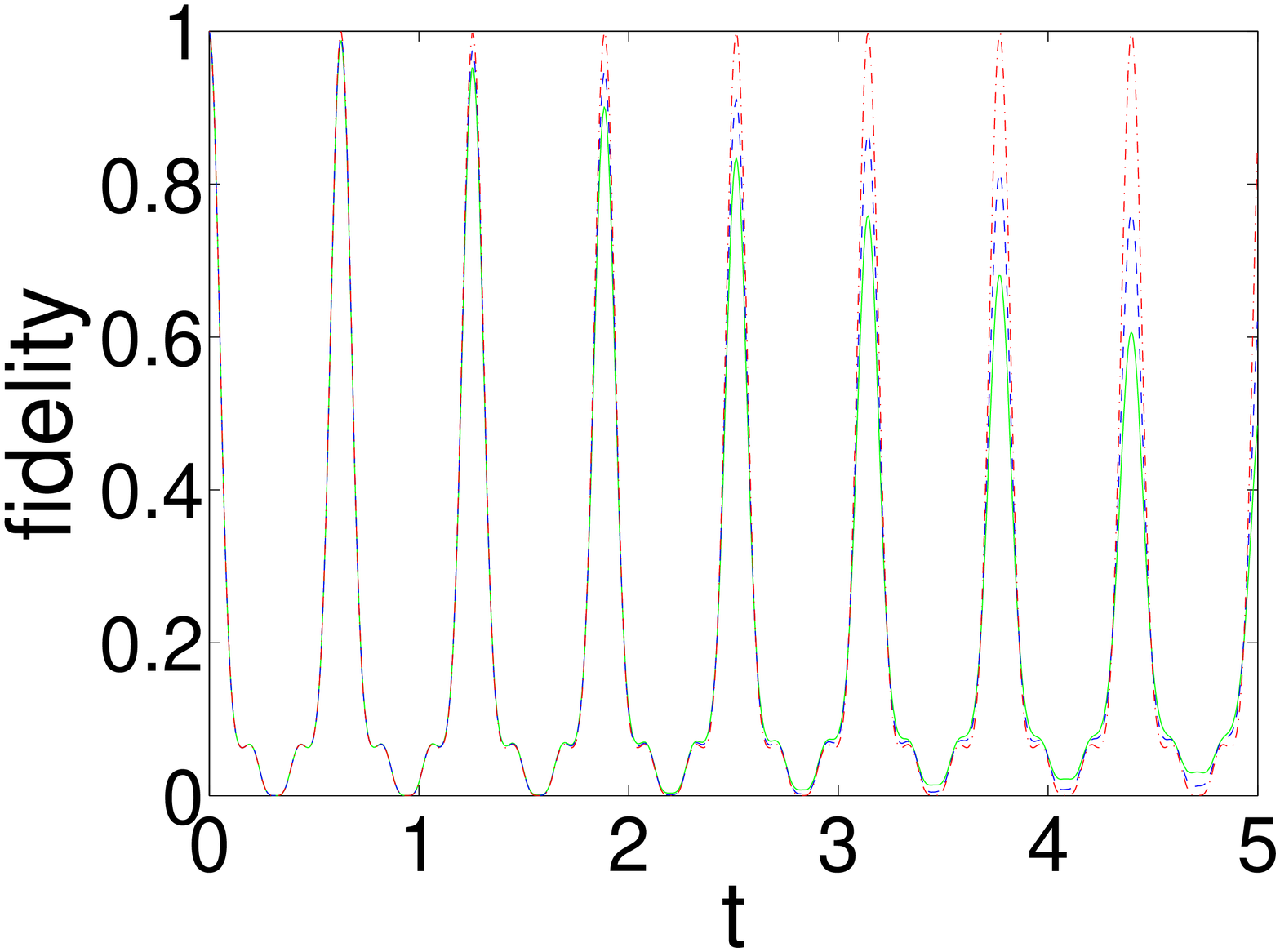}}
  \subfigure[]{
    \includegraphics[width=4.1cm]{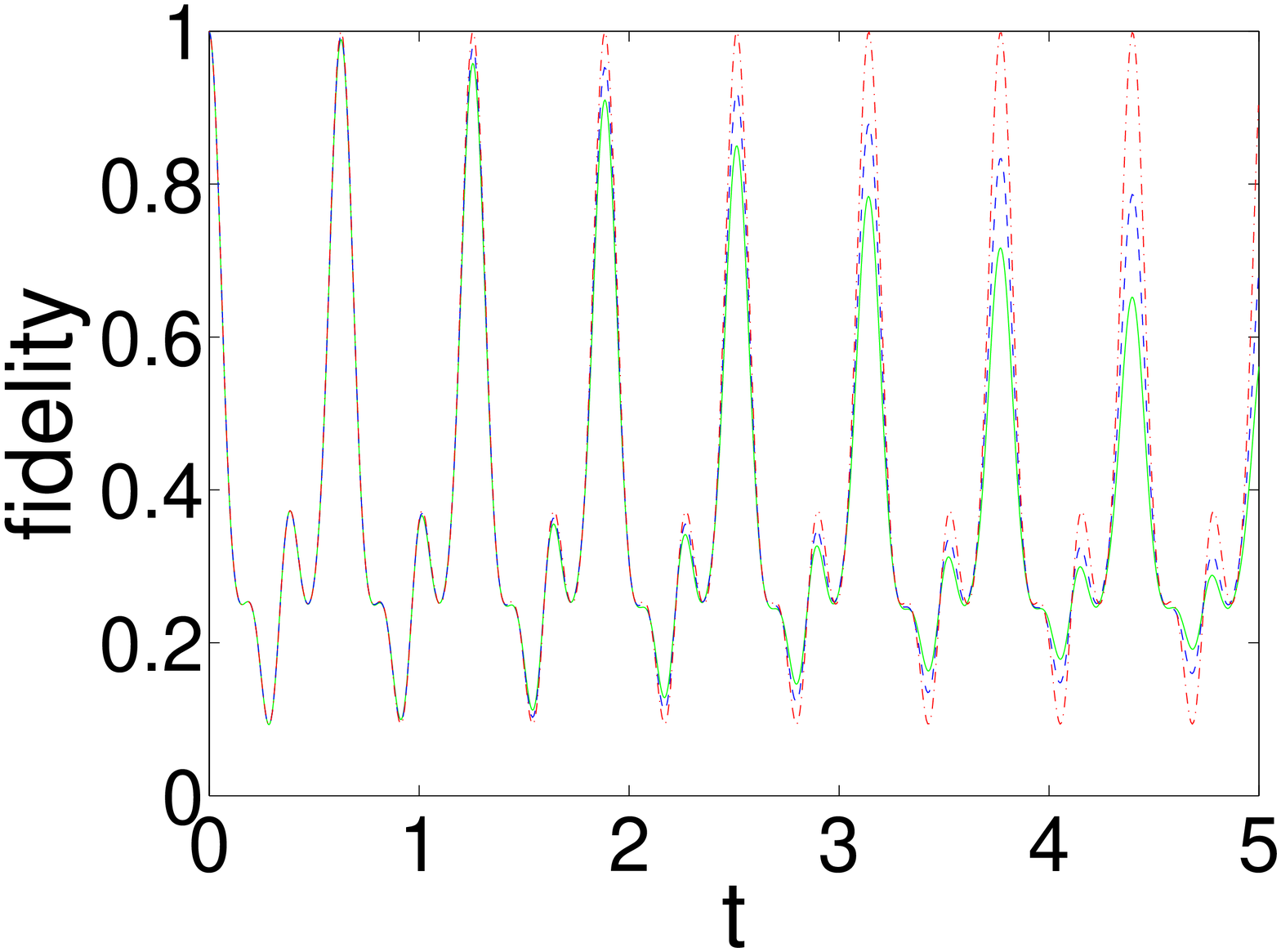}}
  \caption{(Color online) Fidelity-time dependence. Solid green lines for $\theta=\pi/2$, dashed blue lines for $\theta=\pi/4$, and dash-dotted red lines for $\theta=0$. (a) 5-qubit identity gate. (b) 8-qubit Hadamard gate. (c) CZ gate. (d) Z-rotation gate, $\zeta=\pi/8$.}
  \label{rev1}
\end{figure}

The gate fidelity is still defined by equation (\ref{originalGateFid}). We show the calculation outcomes of 5-qubit identity gate, 8-qubit Hadamard gate, Z-rotation gate and controlled-Z gate. We set $g=0.1$, $\epsilon=5$, $T=1$, with different $\theta$ considered (see Fig. \ref{rev1}).

Fidelity curves in Fig. \ref{rev1} show some common pattern. First, all angle $\theta$ results in similar oscillation patterns, with the same oscillation frequency. The reason is transparent. Since the coupling is weak, we are safe to assume that the oscillation frequency is mainly governed by $H_0$ part of the Hamiltonian, which stays independent against the noise type (or equivalently, the parameter $\theta$).

An interesting fact is that phase noise imposes less damage than amplitude noise. According to the figure, $\theta=0$ cases has its peak fidelity near to $1$ even after a long time. This character can be understood by some qualitative reasoning. Since a Z error cannot change the energy of the qubits while the boson environment must add or subtract its boson number by one to impose an error, the total energy of the system is changed by $2\epsilon$. In contrast, an X error changes the energy of the qubits by $2\epsilon$, which compensates the energy change in the boson environment, and keeps the energy of the whole system unchanged. Since X errors does not require the energy of the system to change, it occurs much easier than Z errors.

\section{Cluster-Hamiltonian Creation Scheme}\label{hamiltoniancreation}
A cluster state can also be created by cluster Hamiltonians \cite{Raussendorf2003, Barlett2006}, of which the cluster state is the ground state. The simplest cluster Hamiltonian is 
\begin{equation}\label{Hwithout}
H_{fC}=-J \sum_i K_i,
\end{equation}
where $J>0$, and $K_i$ are the stabilizers of the cluster state:
\begin{equation}
K_i=X_i \prod_{\mathrm{neig}} Z_j.
\end{equation}
The product is over all sites neighboring site $i$. Since the ground state $|g\rangle$ must be a state that
\begin{equation}
K_i|g\rangle=|g\rangle,\forall i,
\end{equation}
we claim that $|g\rangle=|\Psi_C\rangle$. Therefore, cooling this system, we obtain a thermal state close to a cluster state.

To construct an appropriate model for noise, we need to study the energy level of Hamiltonian (\ref{Hwithout}). The eigenstate of $H_{fC}$ is just the common eigenstates of all cluster stabilizers $\{K_i\}$. The first excited state, $|e_1\rangle$, would be the eigenstate with eigenvalue $-1$ of a certain $K_i$ and the eigenvector with eigenvalue $1$ of $K_j,\forall j\neq i$. Its energy is $2J$ higher than the ground state. Following this reasoning, we conclude that the energy gap between neighboring energy levels are all $2J$. Degeneracy of this model can also be deduced. For example, since $|e_1\rangle$ can only have one eigenvalue $-1$, and there are $n$ stabilizers for a certain cluster state, we conclude that the first excited state is N-fold degenerate. Applying the same method, one can calculate the degeneracy of higher excited states. We remark that the ground state is non-degenerate.

\begin{figure}
  \centering
  \includegraphics[width=6.5cm]{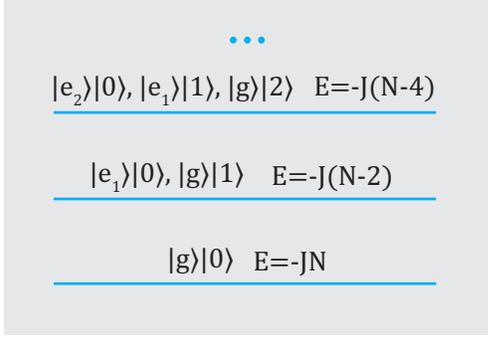}\\
  \caption{(Color online) The energy levels of $H_{C}$, with $g=0$.}\label{energyLevel}
\end{figure}

Now, we can construct a single-mode boson environment, with the mode frequency resonant with the gap between nearest energy levels:
\begin{eqnarray}\label{Hwith}
H_{C}&=& -J \sum_i K_i + 2J a^\dagger a + g ( a^\dagger + a)\nonumber\\
&&\times\sum_{n} (\cos(\theta)\sigma_z^{(n)}-\sin(\theta)\sigma_x^{(n)}).
\end{eqnarray}
The energy level structure of this Hamiltonian, with $g=0$, is plot in Fig. \ref{energyLevel}.
In the figure, $|0\rangle$, $|1\rangle$ and $|2\rangle$ are the boson number states in Fock space. The ground state is a cluster state, $|g\rangle=|\Psi_C\rangle$, coupled with the vacuum $|0\rangle$, and the first excited state is either $|e_1\rangle$ with vacuum, or $|g\rangle$ with one boson. The gap between the nearest energy levels is also $2J$.

The density operator for thermal state is
\begin{equation}
\rho_{th}=e^{-\beta H_C}/\Tr(e^{-\beta H_C}).
\end{equation}
Again, we compute fidelity for gate operations of this density operator. In low temperature, the thermal state mainly comprises the lowest several energy levels. As a result, we again take a cut-off on boson number in our computation.

\begin{figure}
  \centering
  \subfigure[]{
    \includegraphics[width=4.1cm]{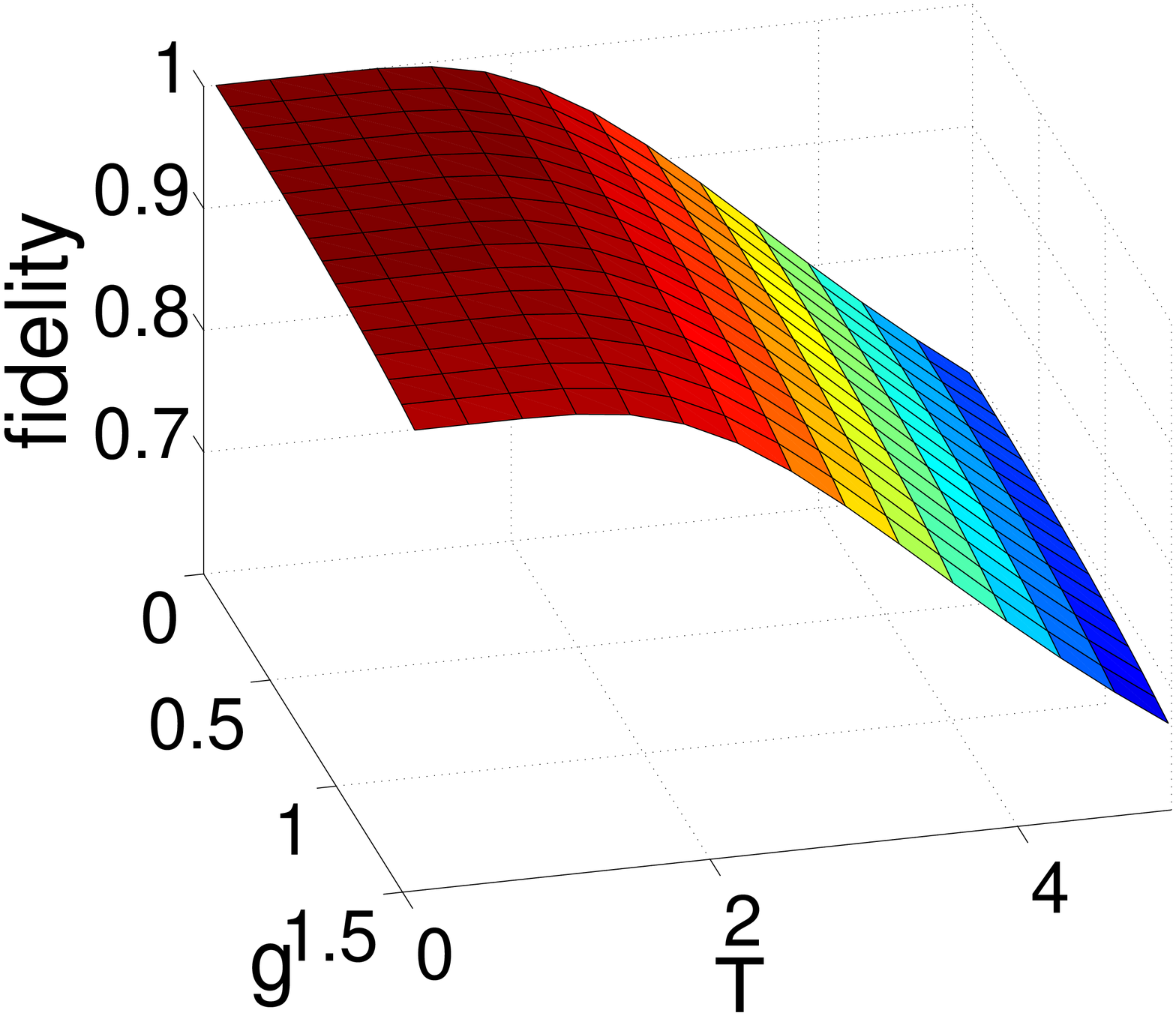}}
  \subfigure[]{
    \includegraphics[width=4.1cm]{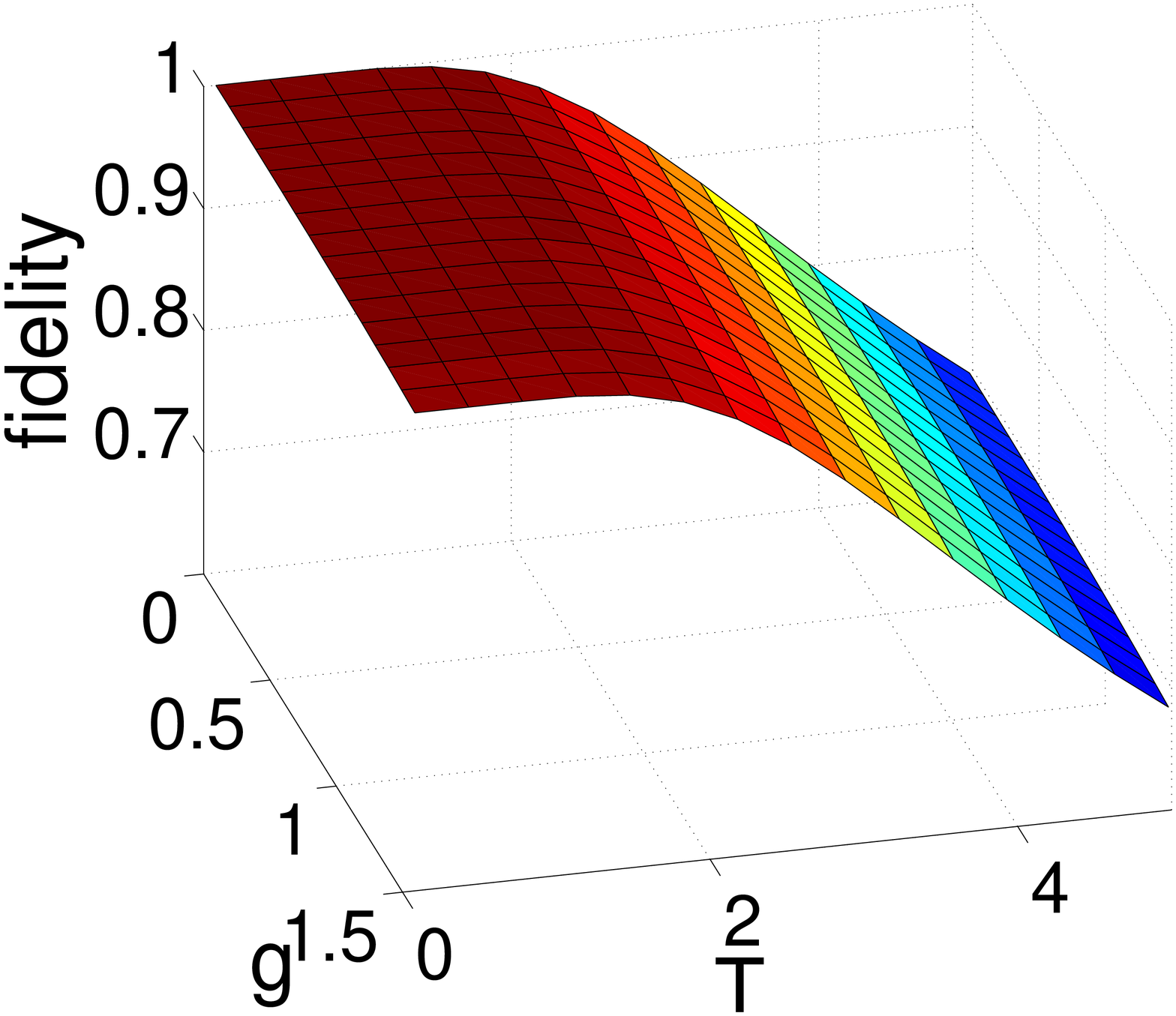}}
\subfigure[]{
    \includegraphics[width=4.1cm]{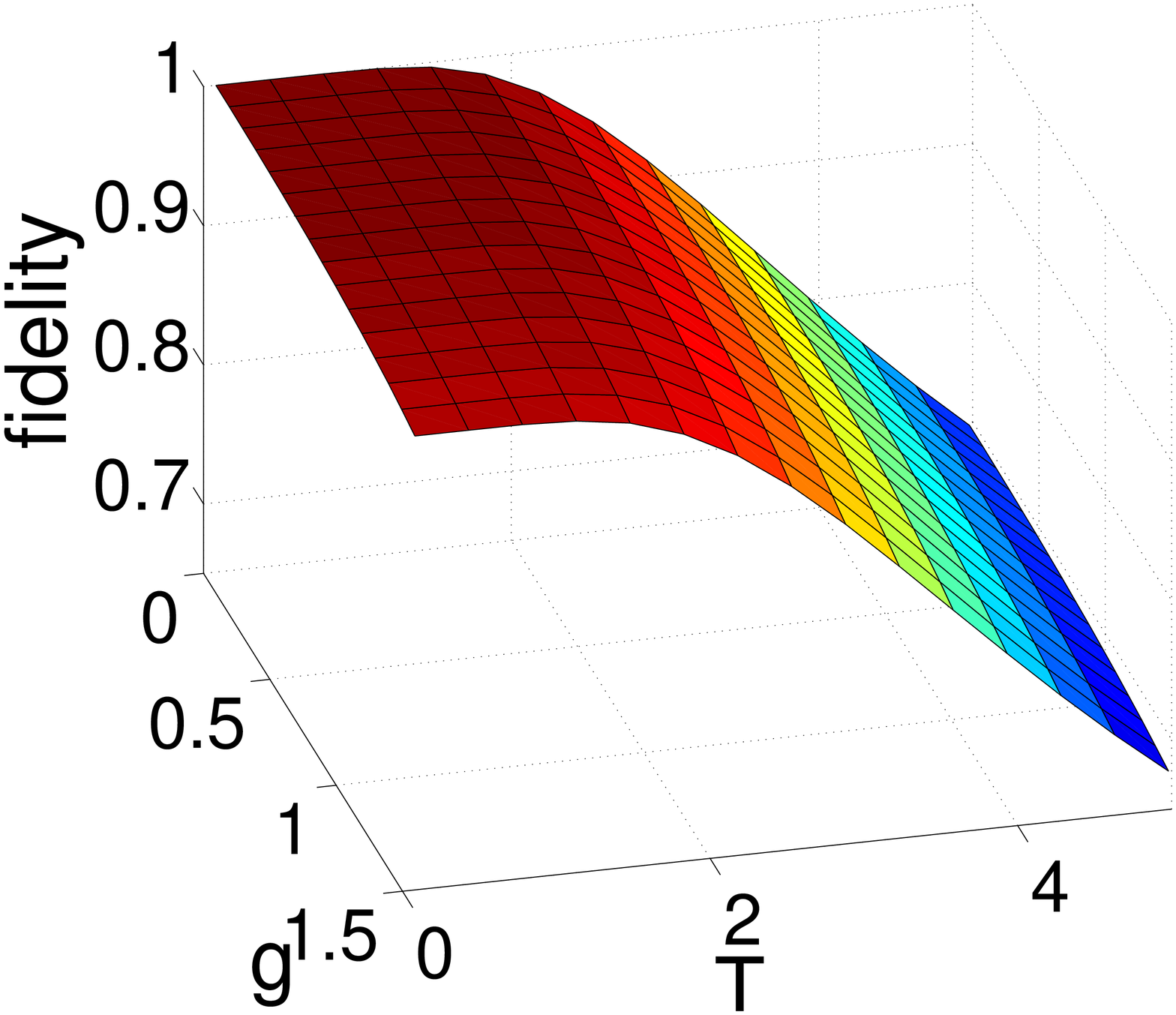}}
  \caption{(Color online) The fidelity of 5-qubit identity gate, depending on $T$ and $g$. (a) $\theta=\pi/2$. (b)$\theta=\pi/4$. (c)$\theta=0$.}
  \label{Fujiirev_id5_3d}
\end{figure}

\begin{figure}
  \centering
  \subfigure[]{
    \includegraphics[width=4.1cm]{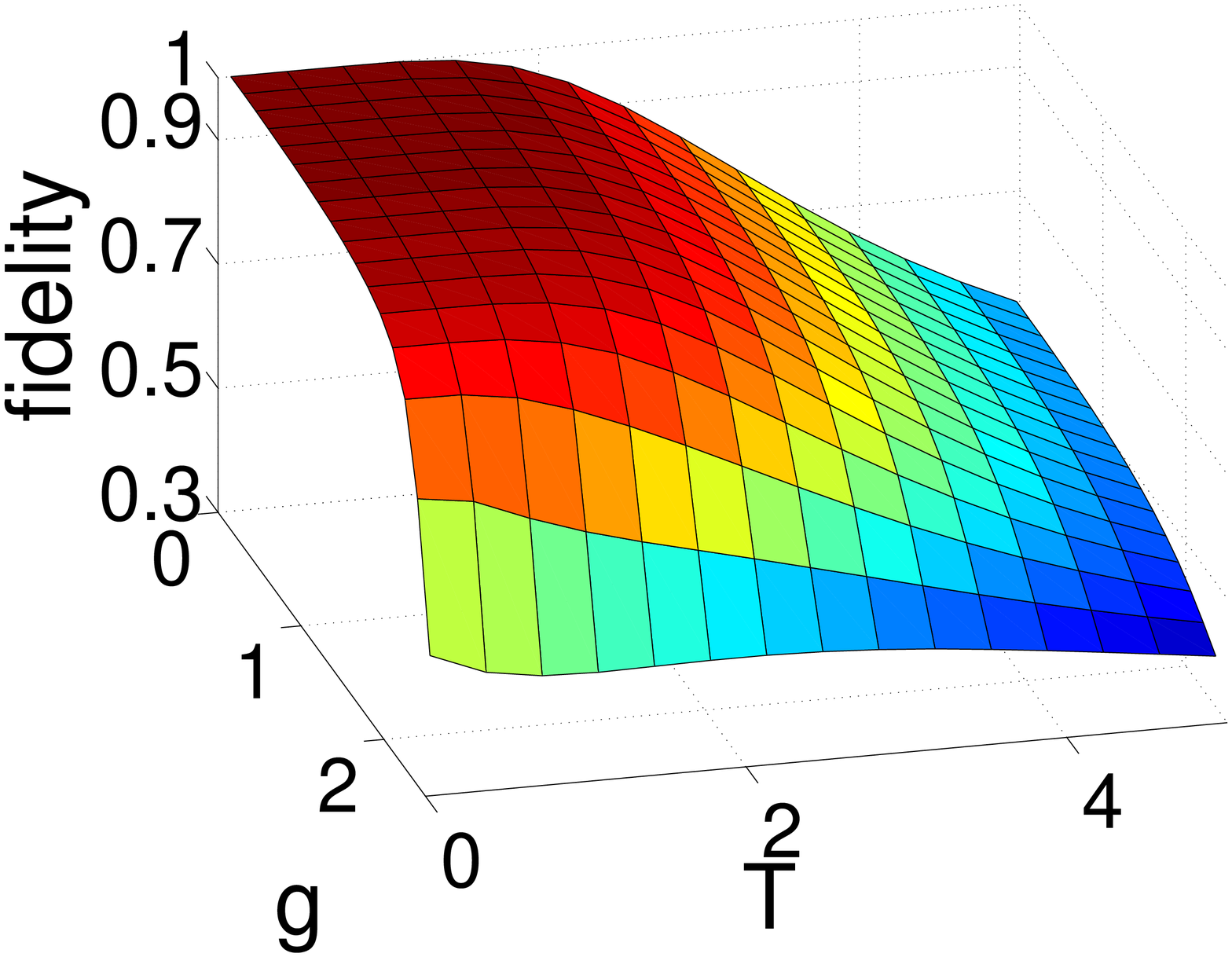}}
  \subfigure[]{
    \includegraphics[width=4.1cm]{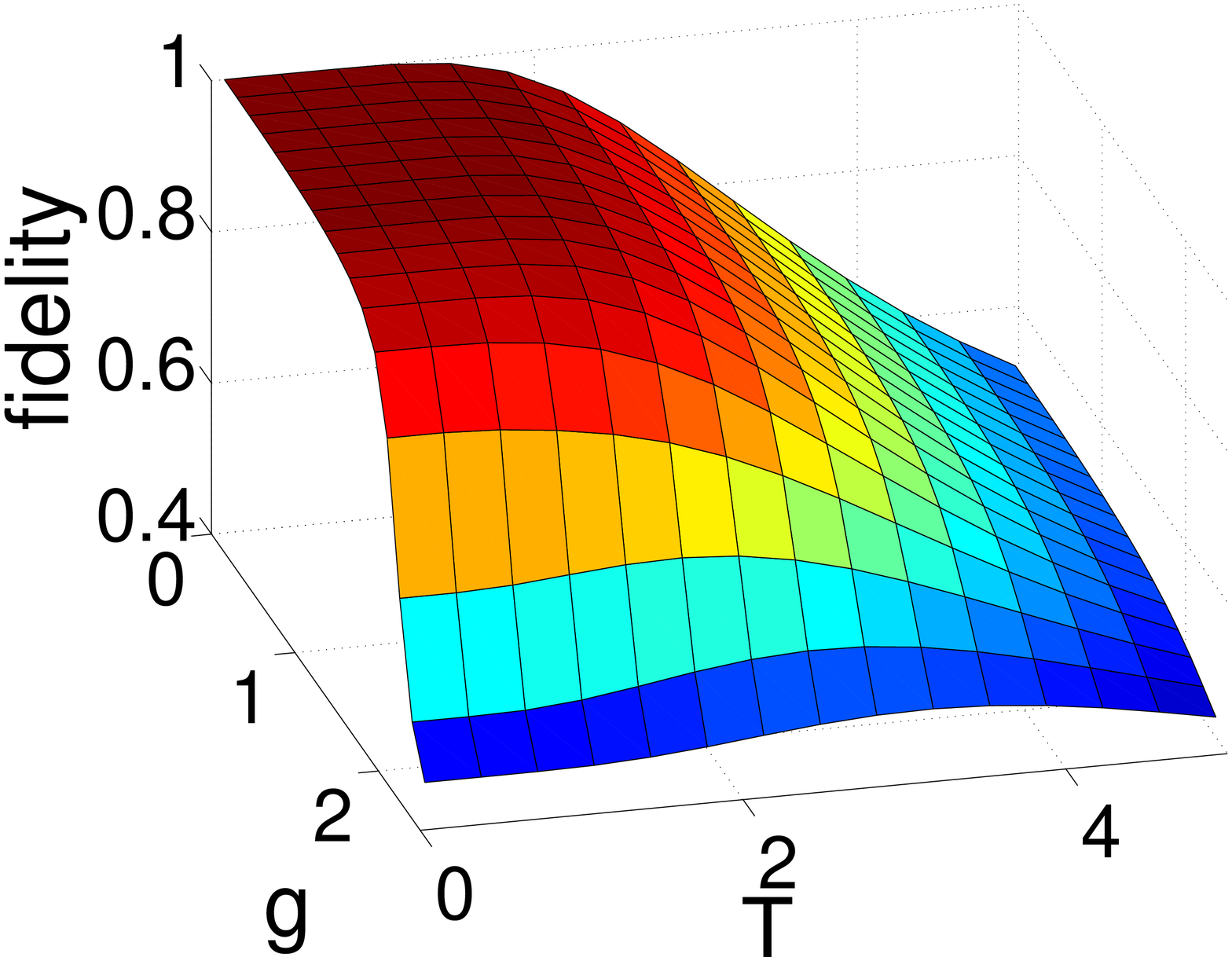}}
\subfigure[]{
    \includegraphics[width=4.1cm]{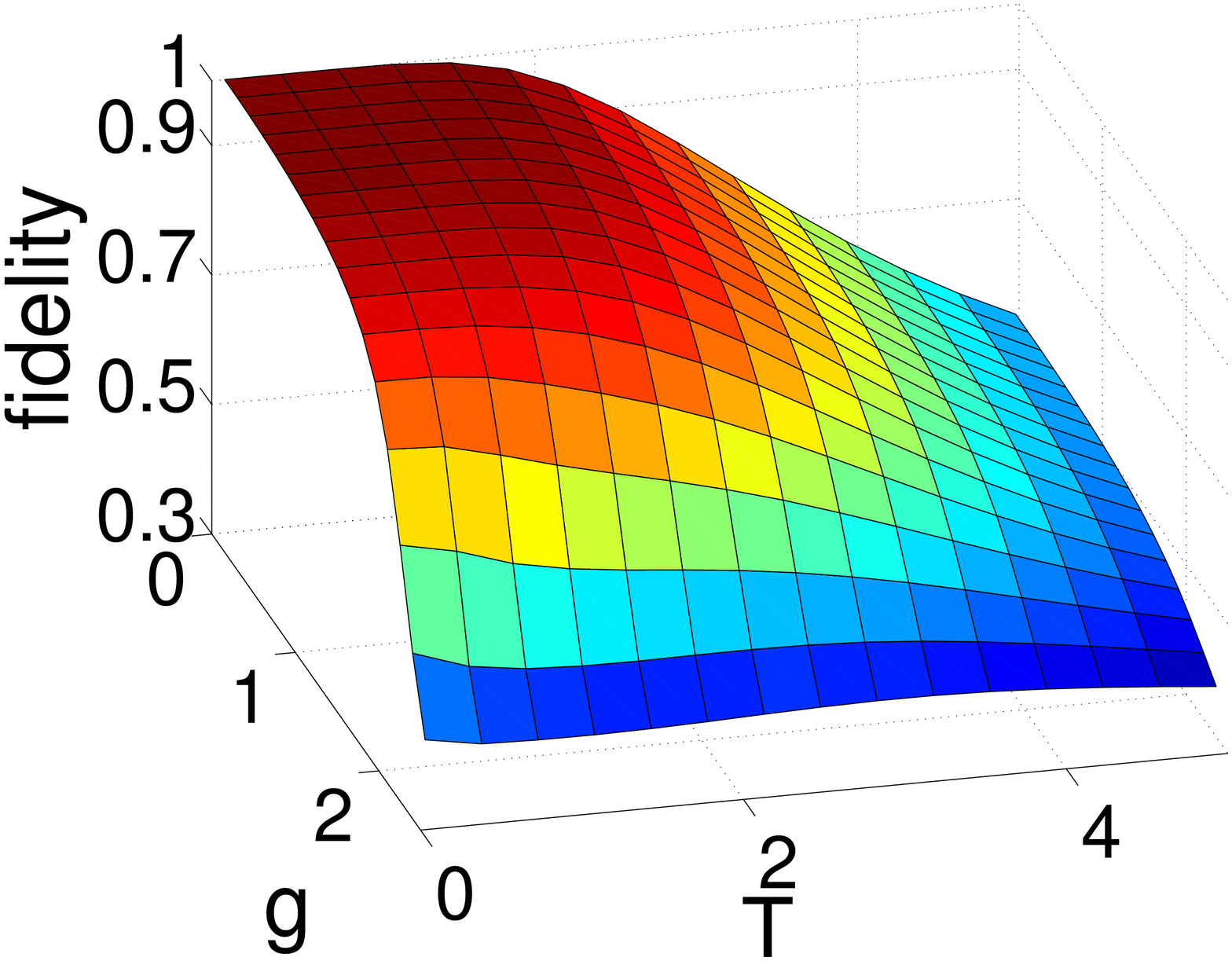}}
  \caption{(Color online) The fidelity of 8-qubit Hadamard gate, depending on $T$ and $g$. (a) $\theta=\pi/2$. (b)$\theta=\pi/4$. (c)$\theta=0$.}
  \label{Fujiirev_har8_3d}
\end{figure}

\begin{figure}
\centering
  \subfigure[]{
    \includegraphics[width=4.1cm]{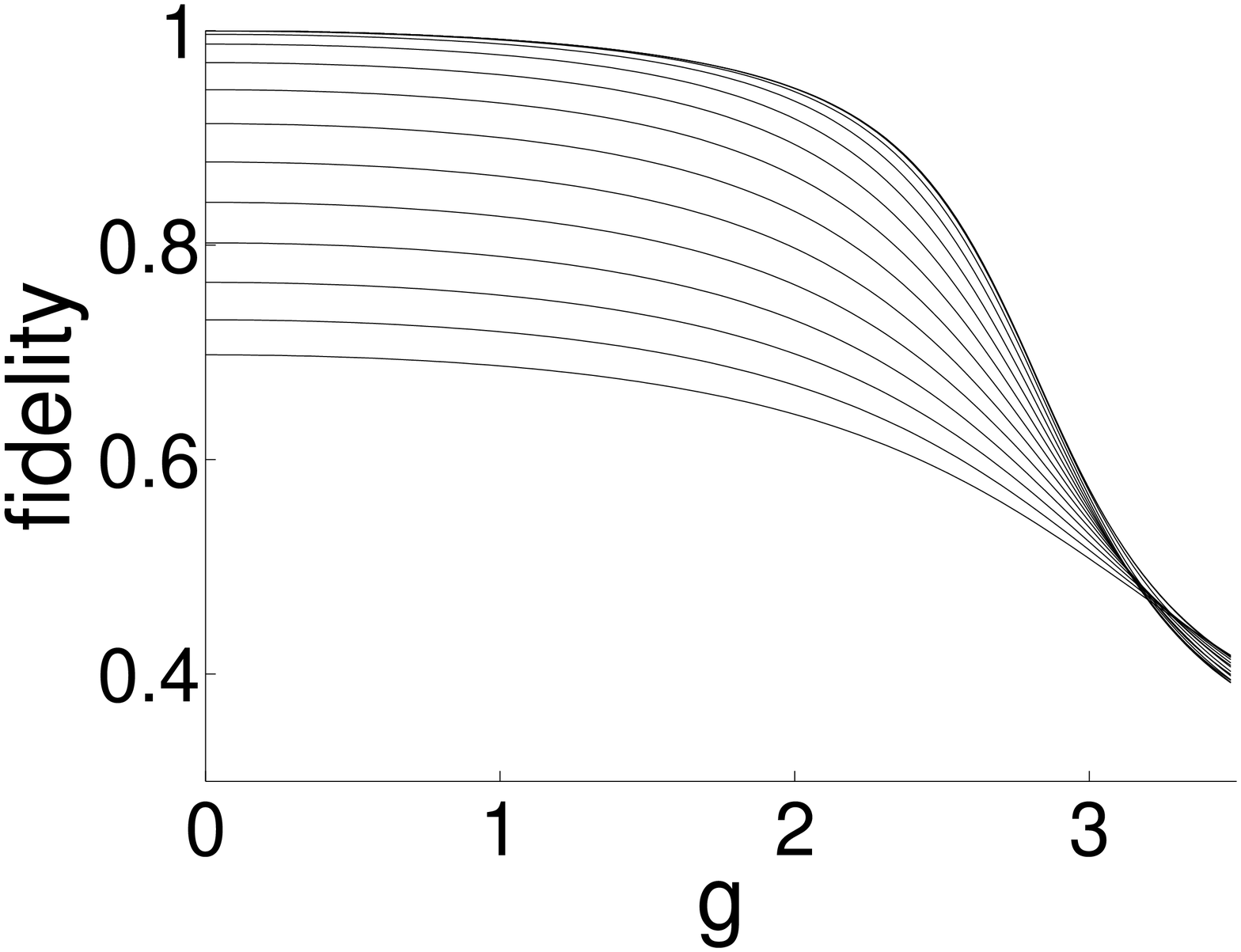}}
  \subfigure[]{
    \includegraphics[width=4.1cm]{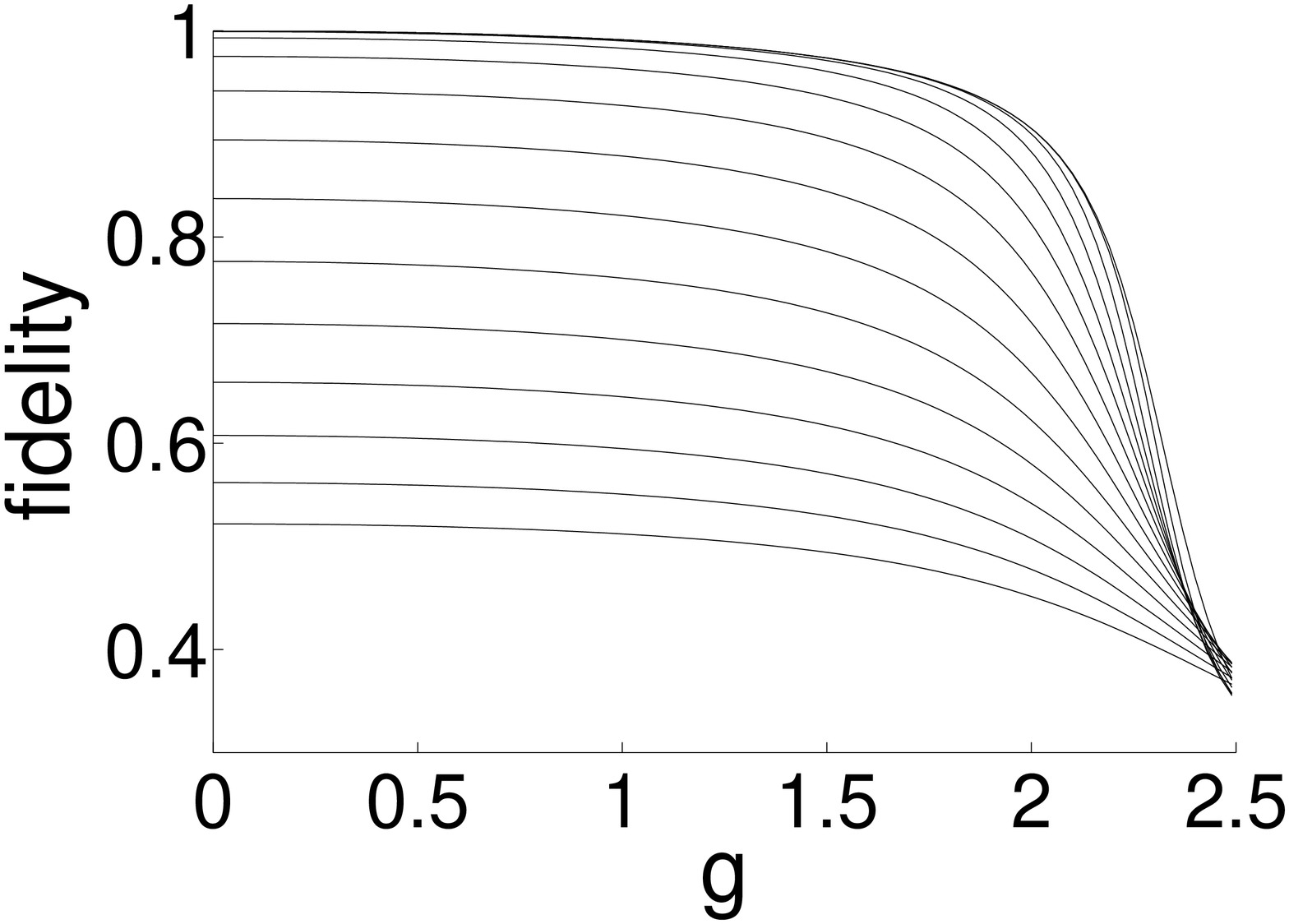}}
  \caption{Fidelity-$g$ dependence. Different lines varies only in temperature. $\theta=\pi/2$. (a) 5-qubit identity gate. (b) 8-qubit Hadamard gate. For small $g$, the fidelity drop is ignorable.}
  \label{fg}
\end{figure}

In Fig. \ref{Fujiirev_id5_3d} and Fig. \ref{Fujiirev_har8_3d}, we plot the fidelity of 5-qubit identity gate and 8-qubit Hadamard gate as a function of temperature $T$ and the coupling coefficient $g$. For clarity, we also present the fidelity-$g$ dependence for different $T$ in one figure, setting $\theta=\pi/2$. See Fig. \ref{fg}.

\begin{figure}
  \centering
  \subfigure[]{
    \includegraphics[width=4.1cm]{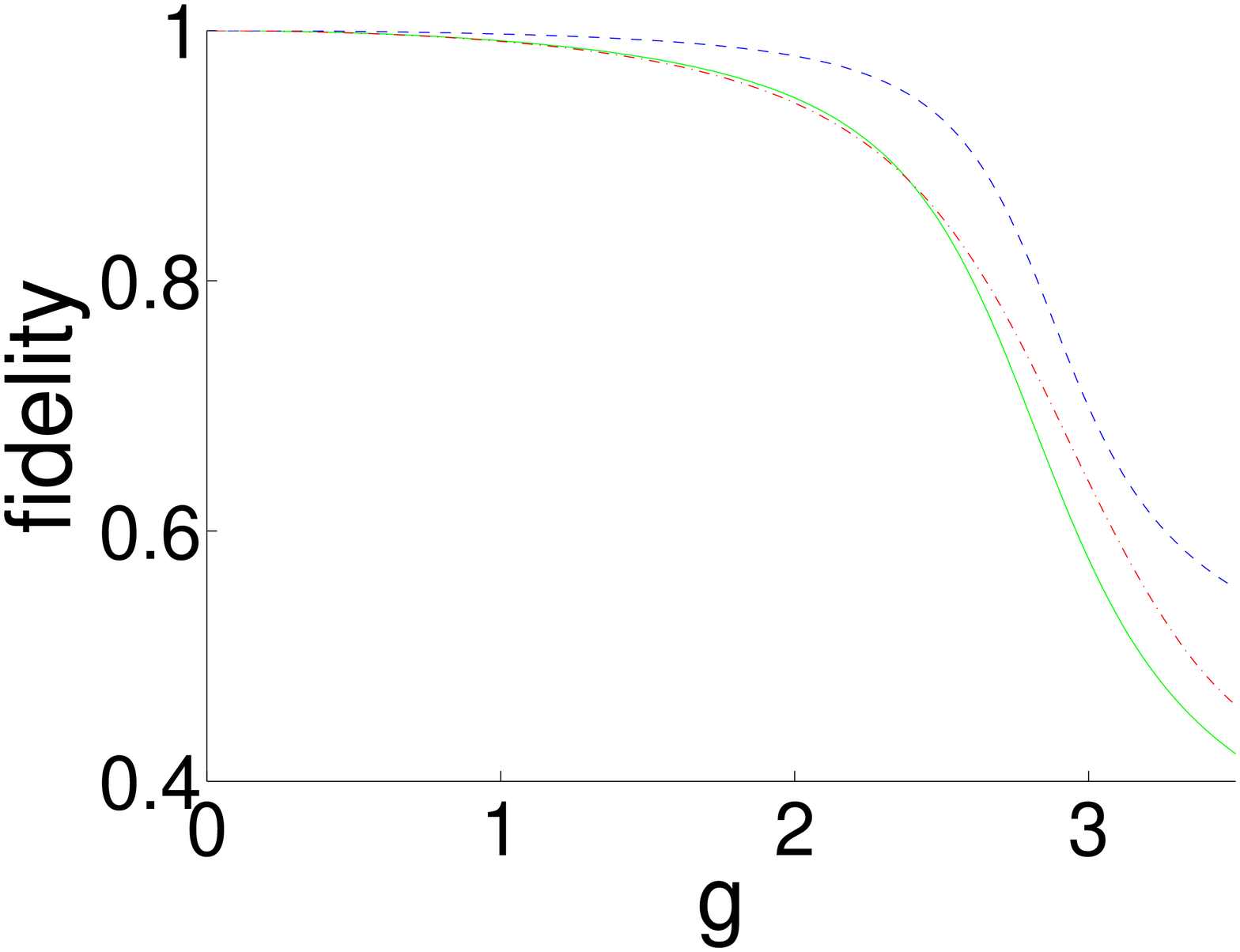}}
  \subfigure[]{
    \includegraphics[width=4.1cm]{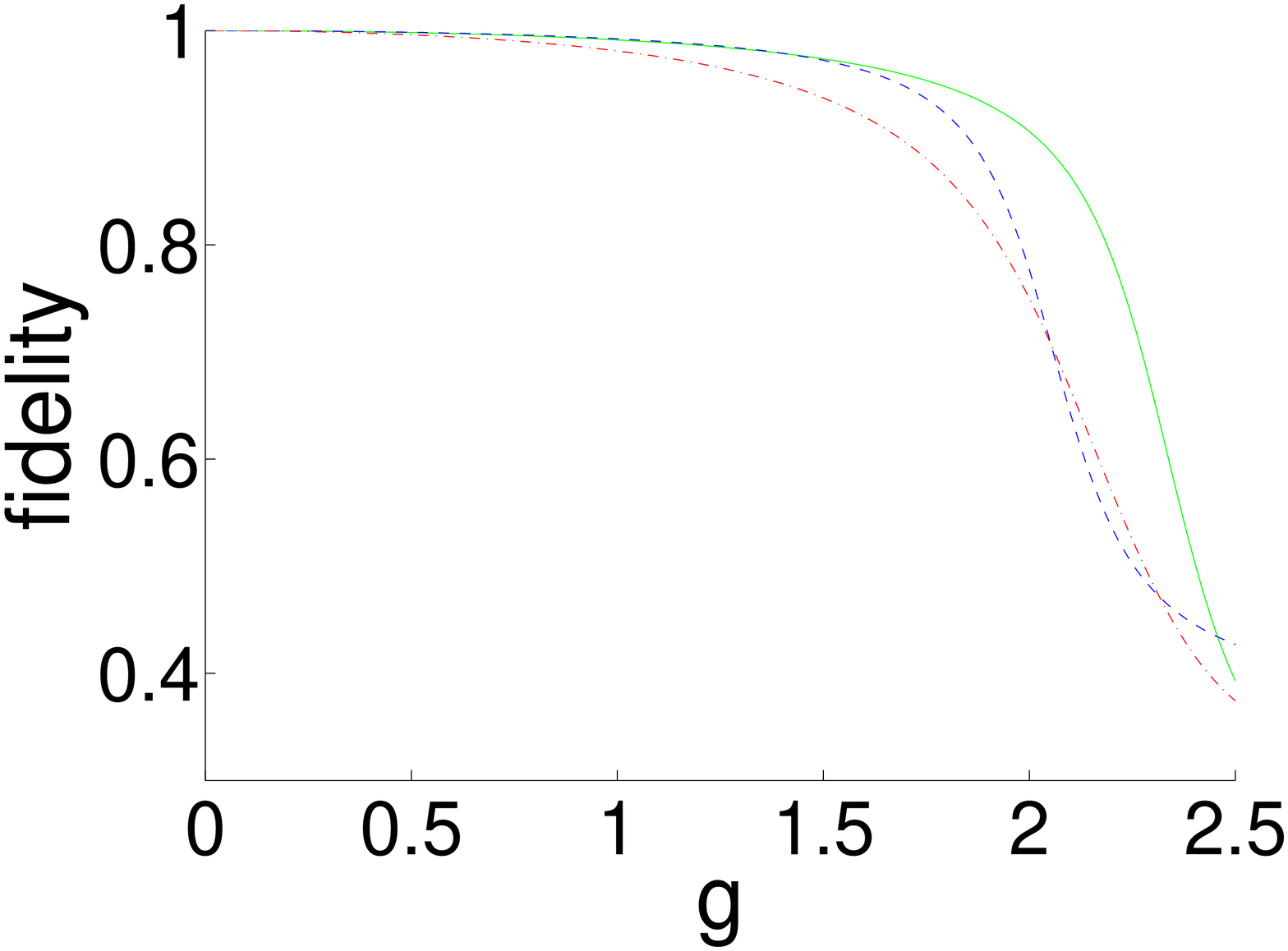}}
  \caption{(Color online) Fidelity-g dependence for different $\theta$. Solid green lines for $\theta=\pi/2$, dashed blue lines for $\theta=\pi/4$, and dash-dotted red lines for $\theta=0$. (a) 5-qubit identity gate. (b) 8-qubit Hadamard gate.}
  \label{fg_theta}
\end{figure}

We find the Hamiltonian creation scheme produces cluster states robust against boson environment, when the coupling is below a certain threshold amount, whatever it is under phase or amplitude noise. The fidelity is close to $1$ as long as $g$ and $T$ is below a critical coupling coefficient value, depending on gate type (see Fig. \ref{fg_theta}). Further, a small $g$ value do not deteriorate the fidelity even when the temperature $T$ is high. This can be observed in Fig. \ref{fg}, where with small $g$, the curves overlap greatly.

When the coupling becomes gradually stronger, a fidelity sudden drop is observed. This character is similar to the fidelity sudden change when $T$ goes high, which is well treated in Ref. \cite{Fujii2013}. Fig. \ref{fg} and Fig. \ref{fg_theta} shows that for 5-qubit identity gate, $g_c\approx2.9$, while for 8-qubit Hadamard gate, $g_c\approx2.4$. Below these critical coupling coefficients, the fidelity does not suffer from the coupling effect. For example, setting $\theta=\pi/4$ and $\epsilon=5$, the fidelity of 5-qubit identity gate at $g=0$, $T=1.83$ is $0.9874$, while at $g=2.4$, $T=1.83$ is $0.9203$, which means fidelity only drops $0.0671$ by an super strong coupling. Actually, $g/J\sim1$, known as ultra strong coupling, requires special design to achieve in experiment \cite{Diaz2010}, would not occur in a quantum computation task. As a result, the fidelity sudden drop does not harass the gate operations evaluated here. However, in a large cluster state, this character may become an issue. Since $g_c$ depends on the cluster size, it may be probable that a large cluster state has a small critical value. In this case, One must calculate the critical coefficient carefully, and restrict the coupling effect below this level in their MBQC systems.

\section{Discussions}\label{discussions}
This part first analyze the difference between gate fidelity and cluster state fidelity. Then, we discuss the collective character of our environment.

One may wonder why gate fidelity is different from the corresponding cluster state fidelity, as the definition is similar to one another.

The first observation to answer this question is that, to get a correct gate teleportation resource state by the procedure shown in Fig. \ref{gateTele}, one does not necessarily need a cluster state. Again we take the Z-rotation gate for example. If there are X errors on both qubit 2 and 4, the resource state can also be correctly prepared:
\begin{eqnarray}
\rho_U&=&\Tr_p\sum_\mathbf{m} B_\mathbf{m}P_\mathbf{m}(X_2\otimes X_4)|\Psi_C\rangle\langle\Psi_C|(X_2\otimes X_4)P_\mathbf{m}B_\mathbf{m}^\dagger\nonumber\\
&=&\Tr_p\sum_\mathbf{m}(X_2\otimes X_4) B_\mathbf{m}P_\mathbf{m}|\Psi_C\rangle\langle\Psi_C|P_\mathbf{m}B_\mathbf{m}^\dagger(X_2\otimes X_4)\nonumber\\
&=&\Tr_p\sum_\mathbf{m} B_\mathbf{m}P_\mathbf{m}|\Psi_C\rangle\langle\Psi_C|P_\mathbf{m}B_\mathbf{m}^\dagger,
\end{eqnarray}
where the relationship $P_\mathbf{m}(X_2\otimes X_4)=(X_2\otimes X_4)P_\mathbf{m}$ holds, since the measurement of qubit 2 and 4 is under $X$ basis. In this section, we denote Pauli X operator as $X$ and Z operator as $Z$. The same reasoning holds when the error is $Z_2\otimes Z_4$. It is thus natural that gate fidelity does not equal to the corresponding cluster state fidelity.

Further, we quantify the difference by the eigenstates of cluster state's stabilizers, since they forms a complete orthonormal basis. We would like to denote the eigenstates as
\begin{equation}
|\psi_i\rangle,\forall i\in \{1,2,\cdots,2^n\},
\end{equation}
with $n$ being the qubit number. We denote the cluster state in it as $|\psi_1\rangle$, satisfying
\begin{equation}
K_j|\psi_1\rangle=1,\forall j\in \{1,2,\cdots,n\}.
\end{equation}
We decompose an arbitrary density matrix as
\begin{equation}
\rho=\sum_{i,j}a_{ij}|\psi_i\rangle\langle\psi_j|.
\end{equation}
To the cluster state fidelity,
\begin{eqnarray}
F = \Tr \left(|\Psi_C\rangle\langle\Psi_C|\rho\right)= \Tr \left(|\psi_1\rangle\langle\psi_1|\rho\right)= a_{11}.
\end{eqnarray}
To the gate fidelity, in contrast,
\begin{eqnarray}
&&F_U = \Tr \left(|\Psi_U\rangle\langle\Psi_U|\rho_U\right)\nonumber\\
&=& \Tr \bigg(|\Psi_U\rangle\langle\Psi_U|\Tr_p\sum_\mathbf{m} B_\mathbf{m}P_\mathbf{m} \rho P_\mathbf{m} {B_\mathbf{m}}^\dagger\bigg)\nonumber\\
&=& \Tr \bigg(|\Psi_U\rangle\langle\Psi_U|\Tr_p\sum_\mathbf{m} B_\mathbf{m}P_\mathbf{m} a_{11}|\psi_1\rangle\langle\psi_1| P_\mathbf{m} {B_\mathbf{m}}^\dagger\bigg)\nonumber\\
&+&\Tr \bigg(|\Psi_U\rangle\langle\Psi_U|\Tr_p\sum_\mathbf{m} B_\mathbf{m}P_\mathbf{m} {\sum_{i,j}}^\prime a_{ij}|\psi_i\rangle\langle\psi_j| P_\mathbf{m} {B_\mathbf{m}}^\dagger\bigg),\nonumber\\
\end{eqnarray}
where the primed sum does not go over $i=j=1$. By the process of gate teleportation, the first part on the right hand side equals to $a_{11}$. The second part, which is none zero generally, results in the difference between gate fidelity and the corresponding cluster state fidelity.

\begin{figure}
  \centering
  \subfigure[]{
    \includegraphics[width=0.23\textwidth]{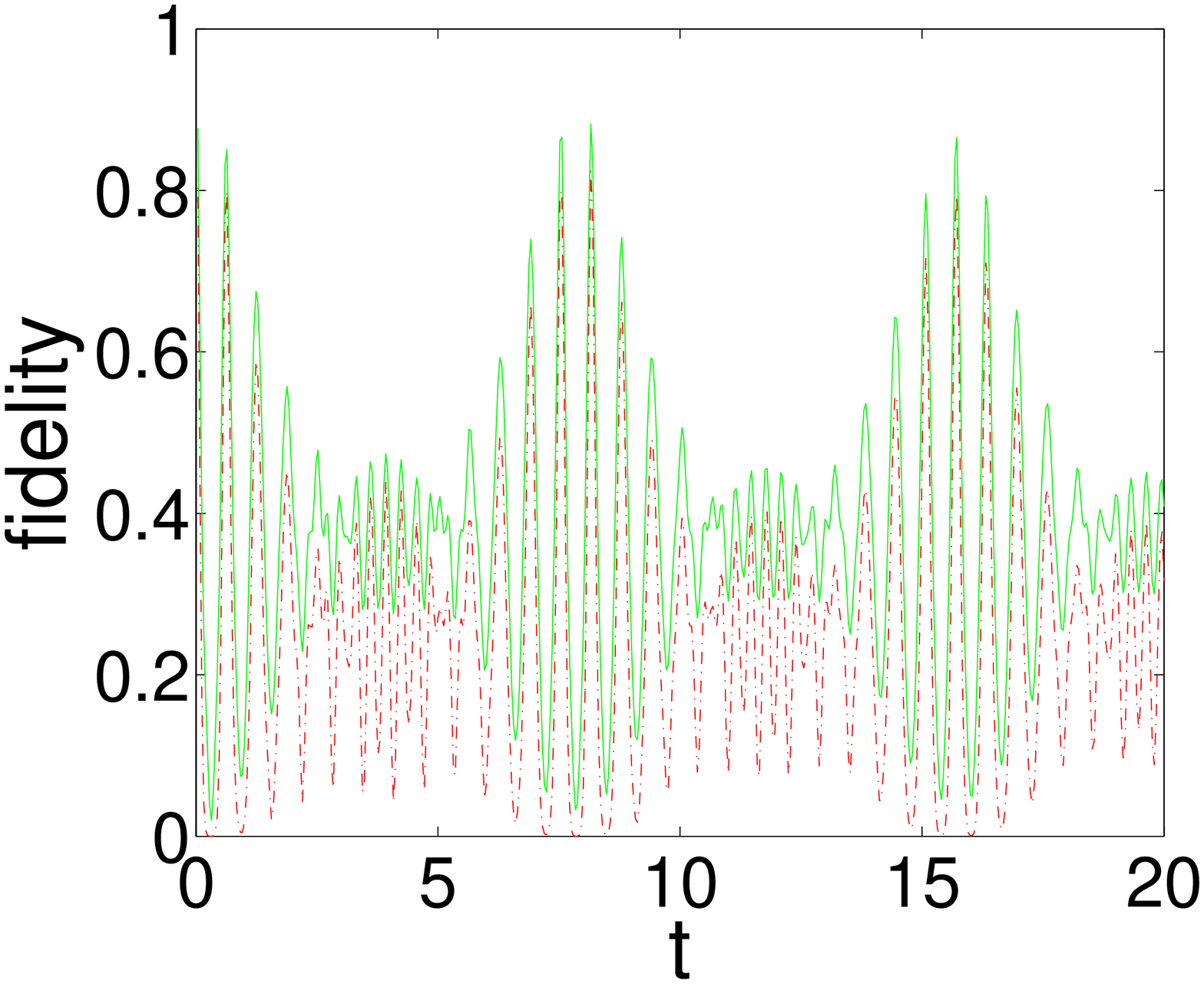}}
  \subfigure[]{
    \includegraphics[width=0.23\textwidth]{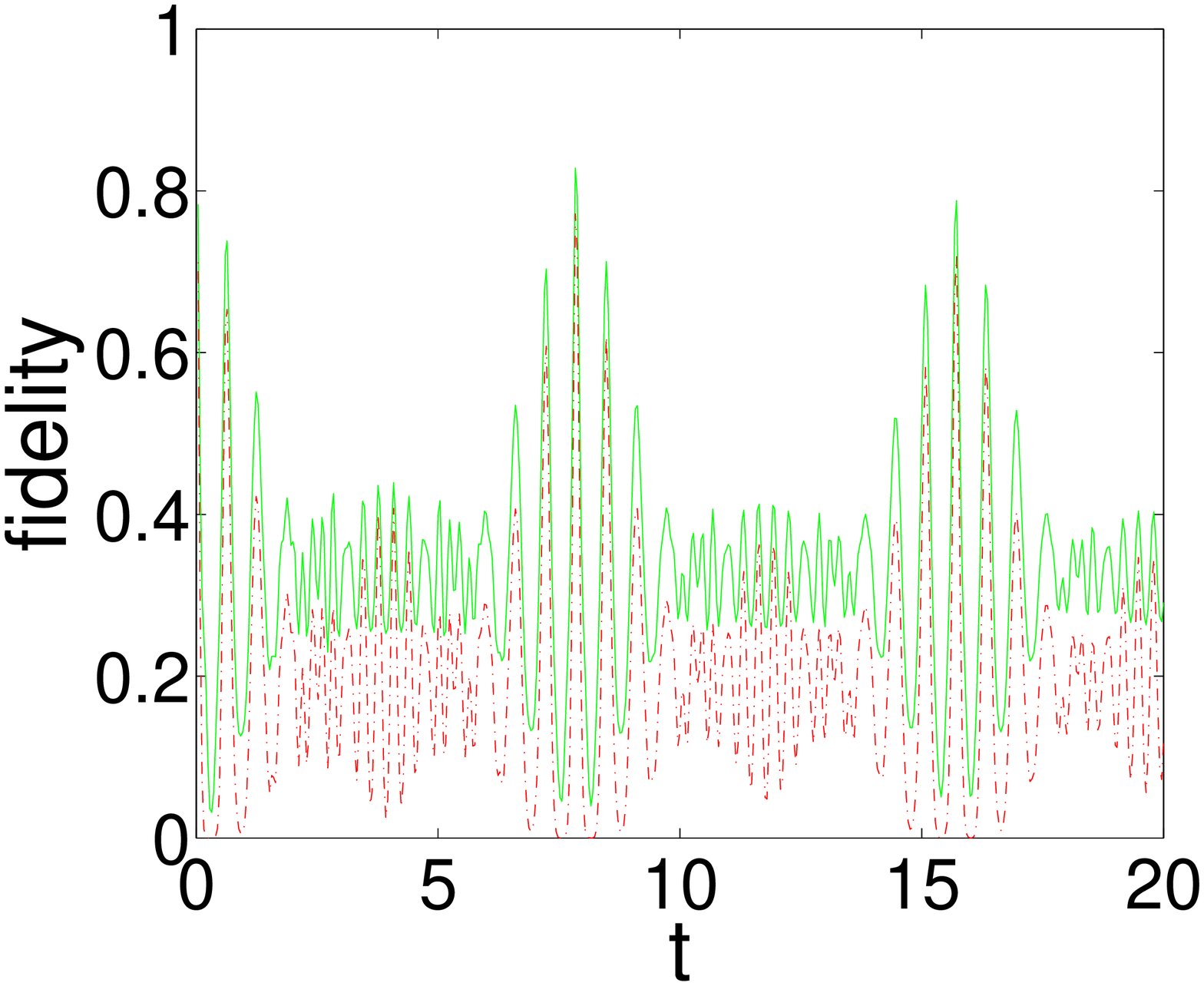}}\\
  \subfigure[]{
    \includegraphics[width=0.23\textwidth]{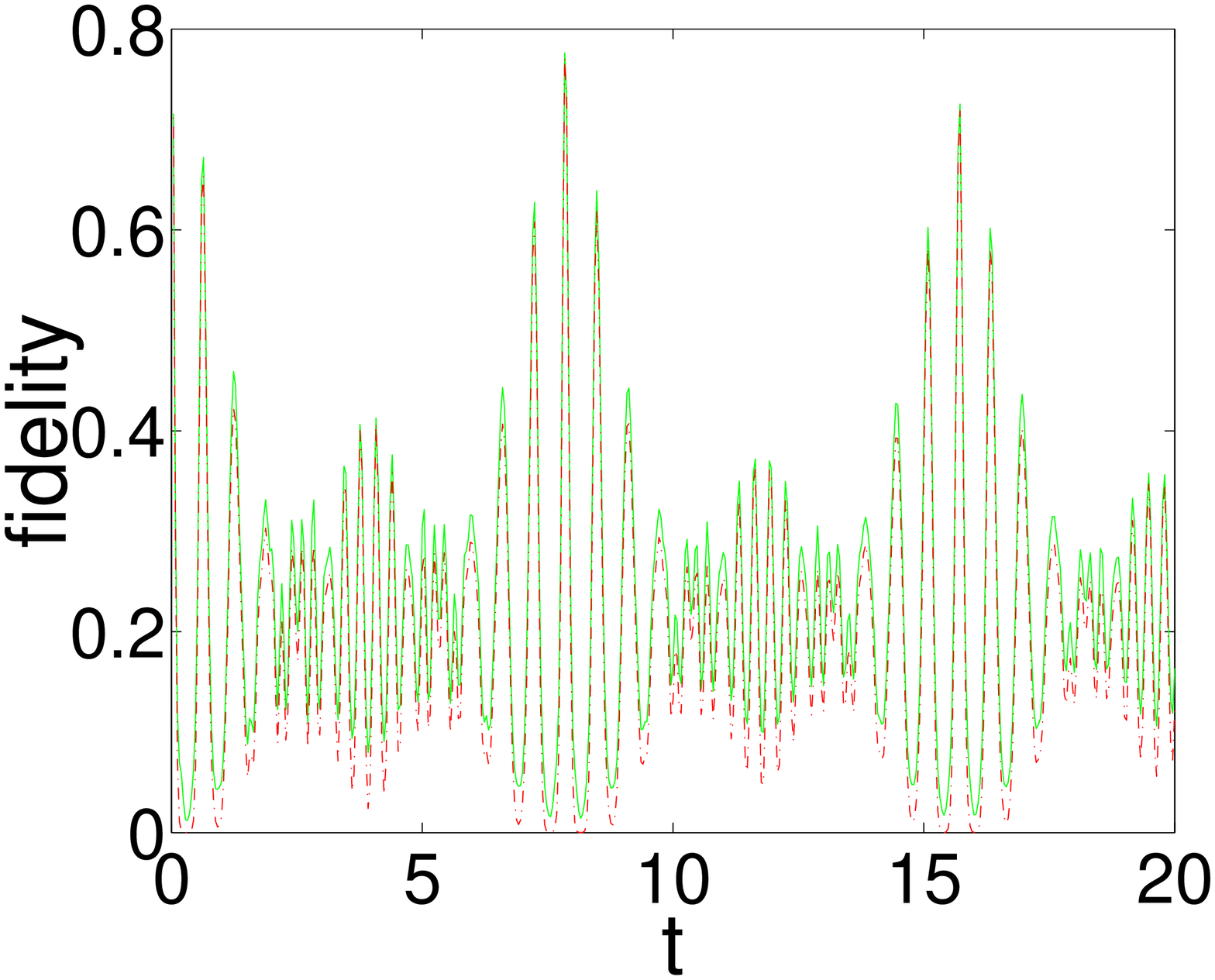}}
  \subfigure[]{
    \includegraphics[width=0.23\textwidth]{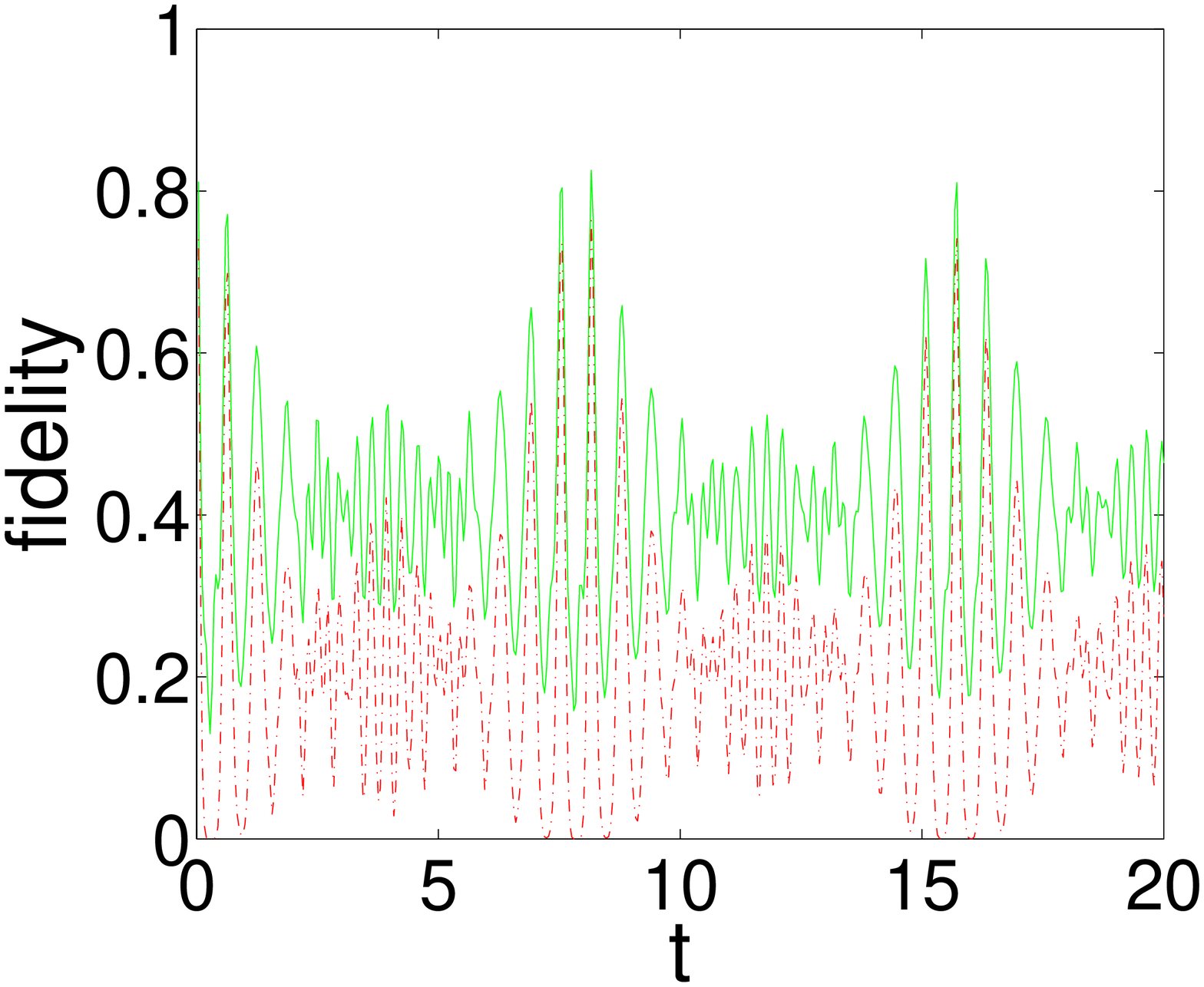}}
  \caption{(Color online) Comparison between gate fidelity (solid green) and the corresponding cluster fidelity (dash-dotted red), with $\epsilon_n=5$. (a) 5-qubit identity gate. (b) 8-qubit Hadamard gate. (c) CZ gate. (d) Z-rotation gate, $\theta=\pi/8$.}
\label{fidelityCompare}
\end{figure}

We plot four fidelity-temperature curves to give a direct impression, see Fig. \ref{fidelityCompare}. The gate fidelity here all goes above the corresponding cluster state fidelity, but generally they are showing the same pattern.

\begin{figure}
  \centering
  \includegraphics[width=0.32\textwidth]{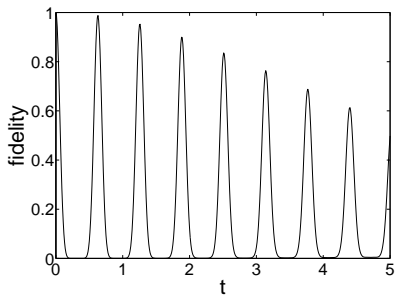}\\
  \caption{A typical oscillation pattern in this setting. The figure is showing the fidelity-time relationship for a 6-qubit linear cluster state.}\label{typical}
\end{figure}

Now, we analyze the collective character of our noise models. Our Hamiltonians, such as Eq. (\ref{exampleHamil}), does not distinguish qubits from each other, and thus keeps invariant when the qubits are permuted. In Ref. \cite{Reina2002}, it is called the ``collective decoherence'' case, to distinguish from the individual decoherence case. The collective character of our Hamiltonian may result in outcomes that is different from the individual case. As an evidence, we evaluate the fidelity drop versus qubit number in a linear cluster state. We use the Hamiltonian (\ref{exampleHamil}). We set $\theta=\pi/2$, $g=0.1$, $\epsilon=5$, and the temperature of the boson environment $T=1$. The typical fidelity oscillation pattern is like Fig. \ref{typical}. Again, the fidelity peaks come at certain time, indifferent to change of the qubit number.
\begin{figure}
  \centering
  \subfigure[]{
    \includegraphics[width=0.23\textwidth]{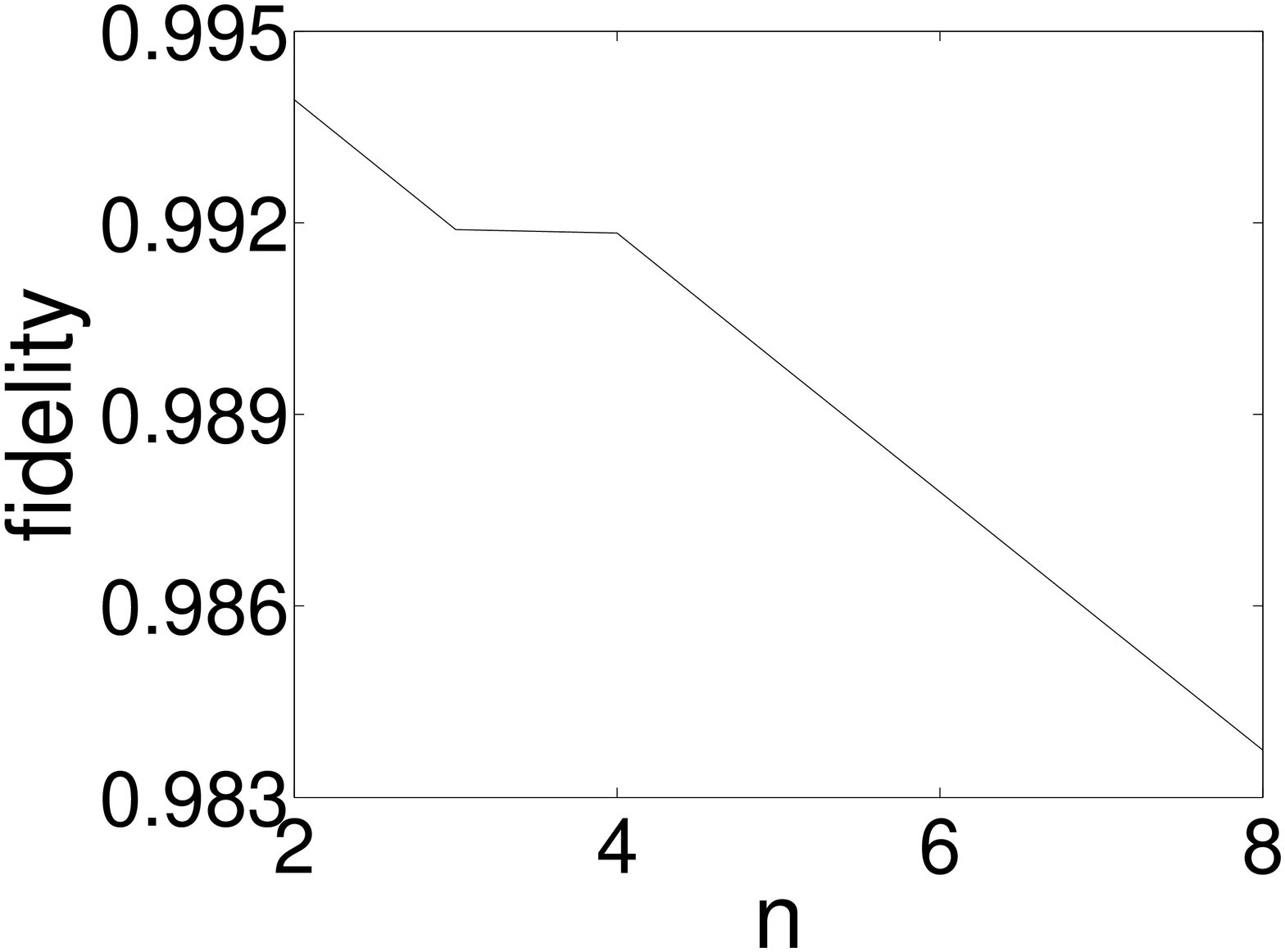}}
  \subfigure[]{
    \includegraphics[width=0.23\textwidth]{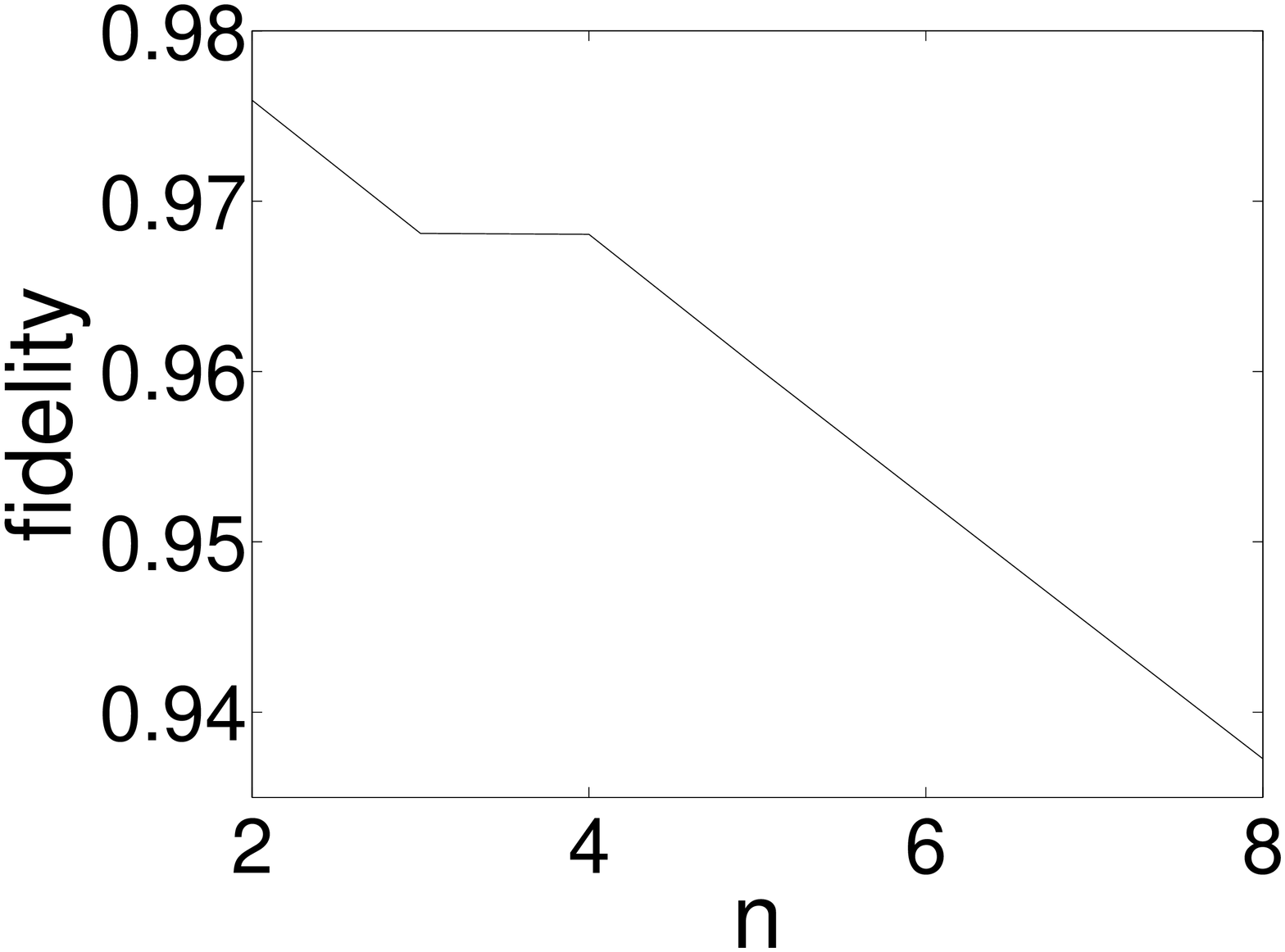}}\\
  \subfigure[]{
    \includegraphics[width=0.23\textwidth]{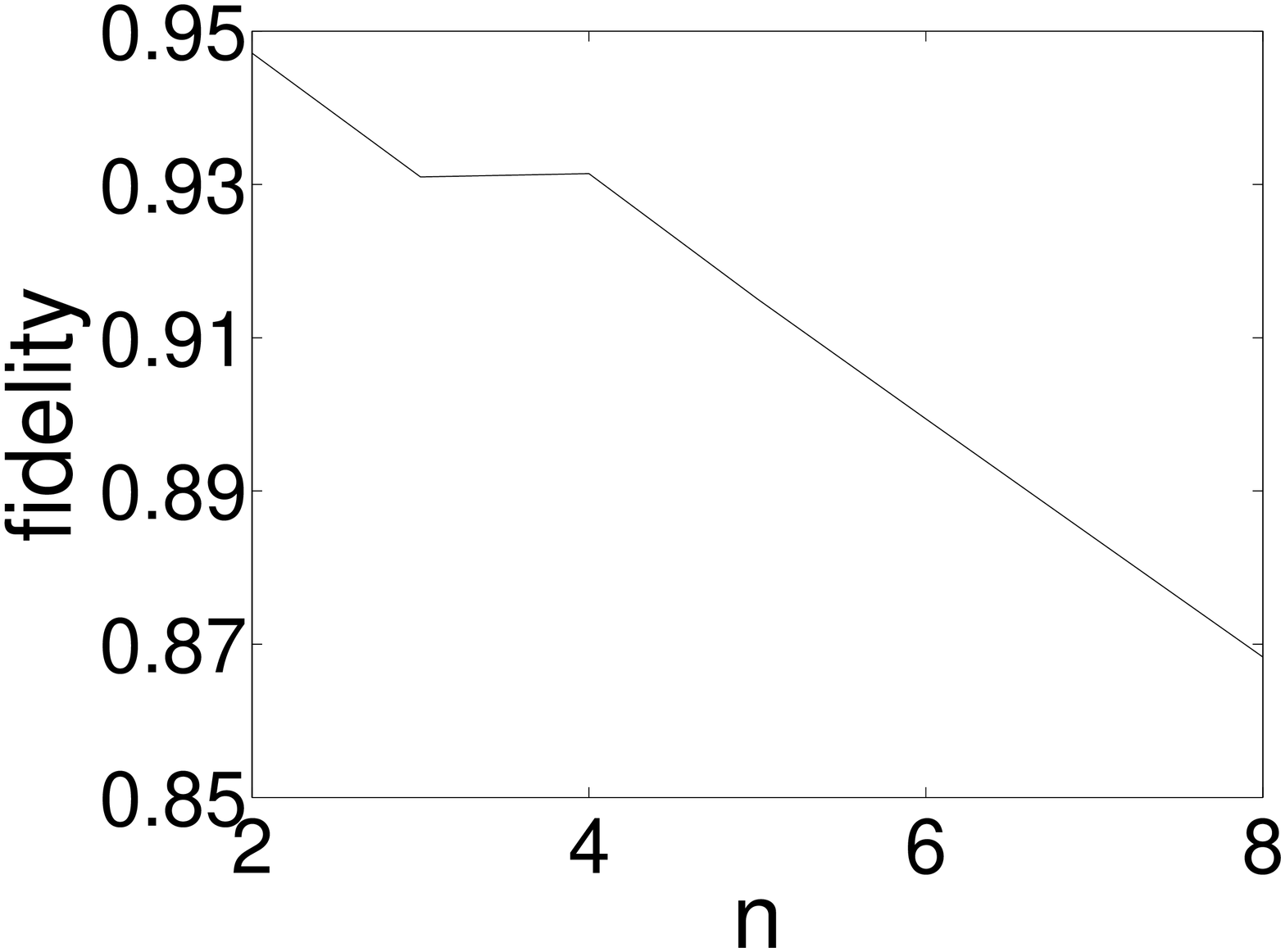}}
  \subfigure[]{
    \includegraphics[width=0.23\textwidth]{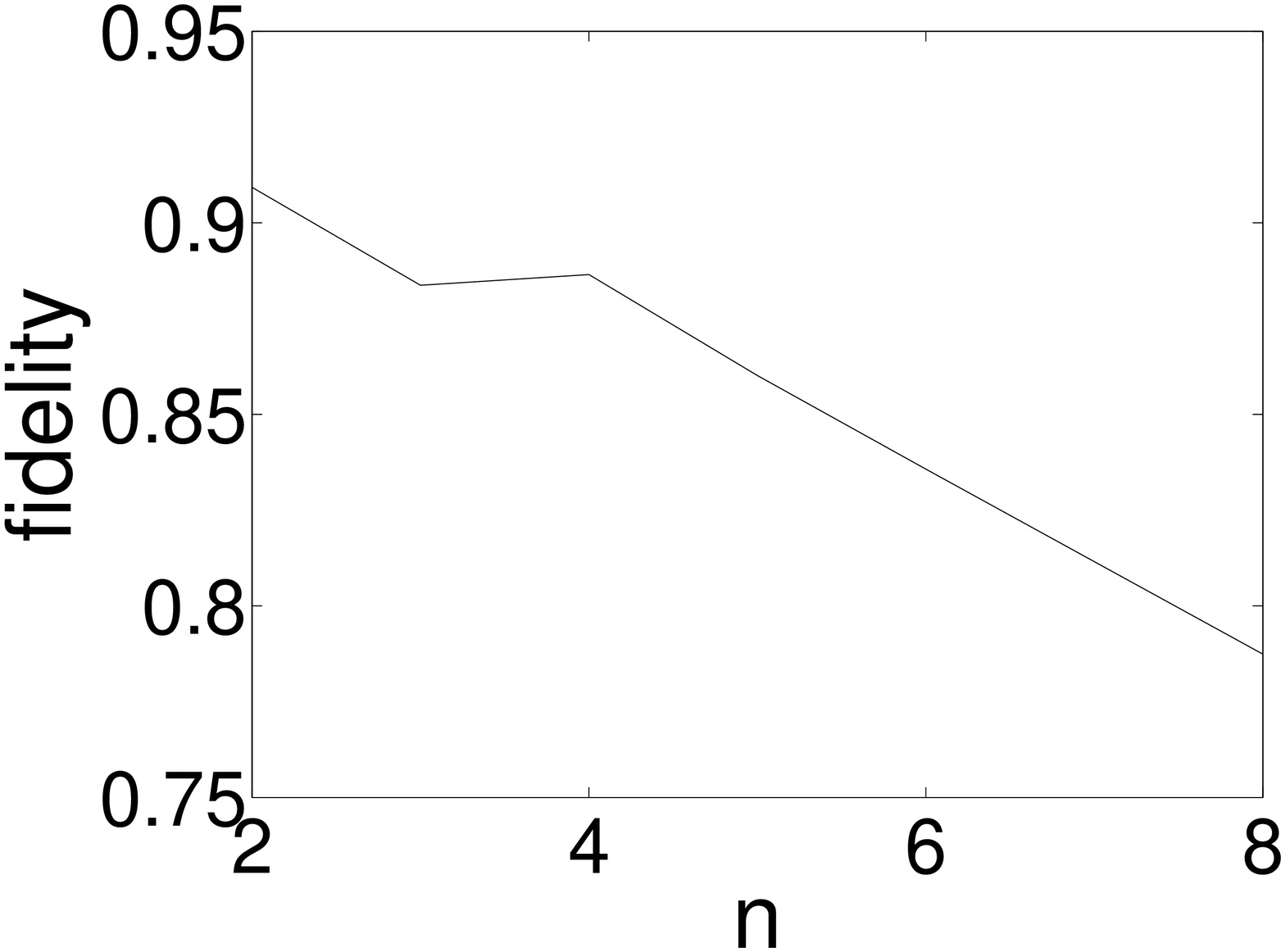}}
  \caption{The cluster fidelity for the first 4 peaks versus qubit number $n$. The fidelity drop turns out to be linear.}
\label{4peaks}
\end{figure}

As is expected in the independent case, the fidelity drops exponentially with the increase of qubits. However, here, the fidelity only drops linearly with the increase of qubit number. This fact may just due to the character of our ``collective'' noise model.

\section{Conclusion}\label{conclusion}
In this paper, we analyze the fidelity of the MBQC scheme when the system is coupled with boson environment. Two specific schemes, the CZ-creation scheme and the Hamiltonian-creation scheme, are studied.

To the CZ-creation scheme, we solve the time evolution of fidelity. We provide the exact solution for the pure phase noise case, and calculate numerically for the generalized noise case. We study four kinds of gate fidelity in detail, and propose two suggestions to enhance its performance under the coupling.

To the Hamiltonian-creation scheme, the fidelity sudden change phenomenon is discovered. Below a certain threshold coupling coefficient value, the damage caused by boson environment is ignorable. The threshold value depends on gate type and cluster size. For individual gates consisting of several qubits, the threshold value is large so one will not worry about the effect caused by the coupling. We conclude that the Hamiltonian-creation method is robust against this kind of noise under a threshold coupling value.

MBQC is a promising quantum computation scheme, and the findings in this paper may be significant to the practical implementation of MBQC systems.

\emph{Acknowledgements:}
This research is supported by `973' program (2010CB922904), NSFC, grants from CAS, NFFTBS (J1030310, J1103205), and the National Undergraduate Innovational Experimentation Program.

\appendix
\section{Derivations of Some Formula}\label{appI}
For completeness, we next present the detailed calculations for two formulas in the main text.  
Calculations here are similar to those in Ref. \cite{Reina2002} and Ref. \cite{Arruda2012}. 
One may also refer to Ref. \cite{Mahan2000} for some part of the calculations.
This part is self-consistent and is presented by
the same style as in the main text.

\subsection{Derivation of the Unitary Evolution Operator}\label{appAi}
The fact that $U_I(t)$ can be solved exactly is due to the pure dephasing character of the Hamiltonian interaction part. In this part of the Appendix, we prove our formula (\ref{ui}) in the main text. Substituting equation (\ref{vi}) into $U_I$, we get
\begin{eqnarray}\label{a3}
U_I(t)&=&\hat{T} \exp\bigg[-i \int_0^t \sum_{n,\mathbf{k}} \sigma_z^{(n)} \nonumber\\
&&\times(g_\mathbf{k} e^{i \omega_k t^\prime} a_\mathbf{k}^\dagger + g_\mathbf{k}^\ast e^{-i \omega_k t^\prime} a_\mathbf{k}) dt^\prime\bigg].
\end{eqnarray}

In order to simplify the expression, we here prove a useful formula
\begin{eqnarray}\label{a4}
&&\hat{T} \exp\left[ \int_0^t dt_1 (\hat{A}(t_1)+\hat{B}(t_1))\right]\nonumber\\
&=&\exp\left[\int_0^t \hat{A}(t_1)dt_1\right]\nonumber\\
&&\times\hat{T}\exp\left\{\int_0^t \left[ \exp\left( -\int_0^{t_1} \hat{A}(t_2) dt_2 \right)\right. \right. \nonumber\\
 &&\left. \left. \times \hat{B}(t_1)  \exp\left(\int_0^{t_1}\hat{A}(t_2) dt_2 \right)\right]dt_1   \right\}.
\end{eqnarray}
First, we denote
$
\hat{X}(t)=\hat{T} \exp[\int_0^t \hat{A}(t_1)+\hat{B}(t_1)dt_1],
$
$
\hat{Y}(t)=\exp[ \int_0^t \hat{A}(t_1)dt_1],
$
and
$
\hat{Z}(t)=\hat{Y}^{-1}(t)\hat{X}(t).
$
We have
\begin{equation}
d\hat{X}/dt=[\hat{A}(t)+\hat{B}(t)]\hat{X}(t),
\end{equation}
\begin{equation}
d\hat{Y}dt=\hat{A}(t)\hat{Y}(t).
\end{equation}
As a result,
\begin{eqnarray}
&&d\hat{Z}/dt\nonumber\\
&=&-\hat{A}(t)\hat{Y}^{-1}(t) \hat{X}(t)+\hat{Y}^{-1}(t)(\hat{A}(t)+\hat{B}(t))\hat{X}(t)\nonumber\\
&=&\hat{Y}^{-1}(t) \hat{B}(t) \hat{X}(t)\nonumber\\
&=&[\hat{Y}^{-1}(t) \hat{B}(t) \hat{Y}(t)]\hat{Z}(t).
\end{eqnarray}
Integrate this equation, we have
\begin{eqnarray}
\hat{Z}(t)=1+\int_0^t [\hat{Y}^{-1}(t_1)\hat{B}(t_1)\hat{Y}(t_1)]\hat{Z}(t_1)dt_1.
\end{eqnarray}
Iterate this equation repeatedly, we get
\begin{eqnarray}
\hat{Z}(t)&=&1+\int_0^t dt_1 \hat{V}(t_1)+\int_0^t dt_1 \int_0^{t_1} dt_2 \hat{V}(t_1)\hat{V}(t_2)+\cdots \nonumber\\
&=&\hat{T}\exp\left[\int_0^t \hat{V}(t_1) dt_1\right].
\end{eqnarray}
where $\hat{V}(t)=\hat{Y}^{-1}(t)\hat{B}(t)\hat{Y}(t)$. We can rewrite this equation as
\begin{eqnarray}
\hat{X}(t)=\hat{Y}(t)\exp\left[\int_0^t\hat{Y}^{-1}(t)\hat{B}(t)\hat{Y}(t)dt_1\right].
\end{eqnarray}
Substituting the expression of $\hat{X}(t)$ and $\hat{Y}(t)$ into it, we prove formula(\ref{a4}).

Now, we evaluate $U_I(t)$ in equation (\ref{a3}). setting
\begin{eqnarray}
\hat{A}(t)&=&-i \sum_{n,\mathbf{k}} \sigma_z^{(n)}g_\mathbf{k}e^{i\omega_k t} a_\mathbf{k}^\dagger,\\
\hat{B}(t)&=&-i \sum_{n,\mathbf{k}} \sigma_z^{(n)}g_\mathbf{k}^\ast e^{-i\omega_k t} a_\mathbf{k},
\end{eqnarray}
(\ref{a4}) becomes
\begin{eqnarray}
U_I(t)&=&\hat{T} \exp\left[ \int_0^t dt_1 (\hat{A}(t_1)+\hat{B}(t_1))\right]\nonumber\\
&=&\exp\left[\int_0^t \hat{A}(t_1)dt_1\right]\nonumber\\
&&\times\hat{T}\exp\left\{\int_0^t \left[ \exp\left( -\int_0^{t_1} \hat{A}(t_2) dt_2 \right)\right. \right. \nonumber\\
 &&\left. \left. \times \hat{B}(t_1)  \exp\left(\int_0^{t_1}\hat{A}(t_2) dt_2 \right)\right]dt_1   \right\}.\label{uii}
\end{eqnarray}
Applying Baker-Hausdorff formula
$
e^{\hat{A}} \hat{B}e^{-\hat{A}}=\hat{B}+[\hat{A},\hat{B}]+[\hat{A},[\hat{A},\hat{B}]]/2!+\cdots
$
and noticing that $[\hat{A},[\hat{A},\hat{B}]]=0$, we conclude
\begin{eqnarray}
&&\exp\left( -\int_0^{t_1} \hat{A}(t_2) dt_2 \right) \hat{B}(t_1)  \exp\left(\int_0^{t_1}\hat{A}(t_2) dt_2 \right)\nonumber\\
&=& B(t_1)-\sum_{n,m,\mathbf{k}}\frac{|g_\mathbf{k}|^2(1-e^{-i\omega_k t_1})}{i\omega_k}\sigma_z^{(n)}\sigma_z^{(m)}.
\end{eqnarray}
As a result, time-ordering is no longer required in the third line of equation (\ref{uii}), and we rewrite it as
\begin{eqnarray}\label{a20}
U_I(t)&=&\exp\left[\int_0^t \hat{A}(t_1)dt_1\right]\exp\bigg\{\int_0^t \bigg[B(t_1)\nonumber\\
&&-\sum_{n,\mathbf{k}}\frac{|g_\mathbf{k}|^2(1-e^{-i\omega_k t_1})}{i\omega_k }\sigma_z^{(n)}\sigma_z^{(m)}\bigg]dt_1   \bigg\}\nonumber\\
&=&\exp\left[\int_0^t \hat{A}(t_1)dt_1 \right]\exp\bigg[\int_0^t B(t_1)dt_1\nonumber\\
&&-\sum_{n,m,\mathbf{k}}\frac{|g_\mathbf{k}|^2(t-\frac{e^{-i\omega_k t}-1}{-i\omega_k})}{i\omega_k }\sigma_z^{(n)}\sigma_z^{(m)}   \bigg]\nonumber\\
&=&\exp\left[\int_0^t \hat{A}(t_1)dt_1\right]\exp\left[\int_0^t B(t_1)dt_1\right]\nonumber\\
&&\times\exp\left[-\sum_{n,m,\mathbf{k}}\frac{|g_\mathbf{k}|^2(t-\frac{e^{-i\omega_k t}-1}{-i\omega_k})}{i\omega_k}\sigma_z^{(n)}\sigma_z^{(m)}   \right].\nonumber\\
\end{eqnarray}
When $[\hat{A},[\hat{A},\hat{B}]]=[\hat{B},[\hat{A},\hat{B}]]=0$, we have $e^{\hat{A}+\hat{B}}=e^{\hat{A}}e^{\hat{B}}e^{-[\hat{A},\hat{B}]/2}$. Applying this formula, we get
\begin{eqnarray}
&&\exp\left[\int_0^t \hat{A}(t_1)dt_1\right]\exp\left[\int_0^t B(t_1)dt_1\right]\nonumber\\
&=&\exp\left[\int_0^t A(t_1)+B(t_1)dt_1\right]\nonumber\\
&&\times\exp\left(\frac{1}{2}\int_0^t dt_1\int_0^t dt_2 [A(t_1),B(t_2)]\right)\nonumber\\
&=&\exp\left[\int_0^t A(t_1)+B(t_1)dt_1\right]\nonumber\\
&&\times\exp\left(\sum_{n,m,\mathbf{k}} \frac{|g_\mathbf{k}|^2}{2{\omega_k}^2} (2-e^{i \omega_k t} -e^{-i \omega_k t})\sigma_z^{(n)}\sigma_z^{(m)}\right).\nonumber\\
\end{eqnarray}
Substituting this into equation (\ref{a20}), we finally get
\begin{eqnarray}
U_I(t)&=&\exp\left[\int_0^t A(t_1)+B(t_1)dt_1\right]\nonumber\\
&&\times\exp\left[i\sum_{n,m,\mathbf{k}} \frac{|g_\mathbf{k}|^2\sigma_z^{(n)}\sigma_z^{(m)}}{{\omega_k}^2} (\omega_k t-\sin\omega_k t)\right],\nonumber\\
\end{eqnarray}
which proves equation (\ref{ui}).

\subsection{Derivation of the Reduced Density Operator of the Qubits}\label{appAii}
In this part of the Appendix, we present the detailed calculation for equation (\ref{rhoi}).

The density operator for the whole system is:
\begin{equation}
\rho_I(t)=U_I(t) \rho^Q_I(0)\otimes \rho^B_I(0) U_I^\dagger(t),
\end{equation}
where the subscript $I$ stands for interaction picture, and the superscript $Q$ and $B$ stands for qubits and boson environment respectively.
What we really care about is the reduced density matrix of the qubits
\begin{equation}
\rho_I^Q(t)=\Tr_B[U_I(t)\rho_I^Q(0)\otimes\rho_I^B(0){U_I}^\dagger(t)].
\end{equation}
We now evaluate each matrix element of $\rho_I^Q(t)$. We define
\begin{equation}
\rho^Q_{I,\{i_n,j_n\}}(t)\equiv\langle i_1,i_2,\cdots, i_N|\rho^Q_I(t)|j_1,j_2,\cdots,j_n\rangle,
\end{equation}
where $N$ is the total number of the qubits, and $i_n=\pm 1$ is the state of the $n$th qubit in the cluster state. We have:
\begin{eqnarray}
\rho_{I,\{i_n,j_n\}}^Q(t)=\Tr[\rho_I^B(0)U_I^{\dagger\{j_n\}}(t)U_I^{\{i_n\}}(t)]\rho_{I,\{i_n,j_n\}}^Q(0),\nonumber\\
\end{eqnarray}
where
\begin{eqnarray}
U_I^{\{i_n\}}(t)&=&\exp\left[i\sum_\mathbf{k} |g_\mathbf{k}|^2 s(\omega_k,t) \sum_{n,m} i_n i_m\right]\nonumber\\
&&\times\exp\bigg\{\sum_{n,\mathbf{k}}[g_\mathbf{k}\varphi_{\omega_k}(t) i_n a_\mathbf{k}^\dagger-g_\mathbf{k}^\ast\varphi_{\omega_k}^\ast(t) i_n a_\mathbf{k}]\bigg\}\nonumber\\
\end{eqnarray}
satisfies
\begin{equation}
U_I(t)|\{i_n\}\rangle=U_I^{\{i_n\}}(t)|\{i_n\}\rangle.
\end{equation}
Explicit calculation reveals
\begin{eqnarray}
&&U_I^{\dagger\{j_n\}}(t)U_I^{\{i_n\}}(t)\nonumber\\
&=&\exp\left[i \sum_\mathbf{k} |g_k|^2 s(\omega_k,t)\sum_{n,m}(i_ni_m-j_nj_m)\right]\nonumber\\
&&\times\exp\left[\sum_{n,\mathbf{k}}[{g_\mathbf{k}}^\ast {\varphi_{\omega_k}}^\ast(t) j_n a_\mathbf{k} -g_\mathbf{k} \varphi_{\omega_k} (t) j_n {a_\mathbf{k}}^\dagger \right]\nonumber\\
&&\times\exp\left[\sum_{n,\mathbf{k}}[g_\mathbf{k} \varphi_{\omega_k} (t) i_n {a_\mathbf{k}}^\dagger -{g_\mathbf{k}}^\ast {\varphi_{\omega_k}}^\ast(t) i_n a_\mathbf{k} \right]\nonumber\\
&=&\exp\left[i \sum_\mathbf{k} |g_k|^2 s(\omega_k,t)\sum_{n,m}(i_ni_m-j_nj_m)\right]\nonumber\\
&&\times\exp\bigg[\sum_{n,\mathbf{k}}[{g_\mathbf{k}}^\ast {\varphi_{\omega_k}}^\ast(t) (j_n-i_n) a_\mathbf{k} \nonumber\\
&&-g_\mathbf{k} \varphi_{\omega_k} (t) (j_n-i_n) {a_\mathbf{k}}^\dagger ]\bigg].
\end{eqnarray}
We here again use the fact that $e^{\hat{A}+\hat{B}}=e^{\hat{A}}e^{\hat{B}}e^{-[\hat{A},\hat{B}]/2}$ when $[\hat{A},[\hat{A},\hat{B}]]=[\hat{B},[\hat{A},\hat{B}]]=0$.
Substituting equation (\ref{rhob}), the initial density operator of the boson environment, into it, we have
\begin{eqnarray}
&&\Tr_B\left[\rho^B_I(0)\exp\left\{\sum_\mathbf{k}(\phi_\mathbf{k} b_\mathbf{k}^\dagger-\phi_\mathbf{k}^\ast b_\mathbf{k})\right\}\right]\nonumber\\
&=&\prod_\mathbf{k}\exp\bigg[-|g_\mathbf{k}|^2\frac{1-\cos(\omega_k t)}{ \omega_k^2} \coth\left(\frac{ \omega_k }{2 k_B T}\right)\nonumber\\
&&\times\sum_{m,n} (i_m-j_m)(i_n-j_n)\bigg],
\end{eqnarray}
where
\begin{equation}
\phi_\mathbf{k}\equiv g_\mathbf{k} \phi_{\omega_k}(t)\sum_n (i_n-j_n).
\end{equation}
This equation leads to
\begin{eqnarray}
&&\rho_{I,\{i_n,j_n\}}^Q(t)\nonumber\\
&=&\exp\bigg[-\sum_{\mathbf{k},m,n}|g_\mathbf{k}|^2c(\omega_k,t)\nonumber\\
&&\times\coth\left(\frac{\omega_k}{2k_B T}\right)(i_m-j_m)(i_n-j_n)\bigg]\nonumber\\
&&\times\exp\left[i\sum_{\mathbf{k},m,n}|g_\mathbf{k}|^2s(\omega_k,t)(i_mi_n-j_mj_n)\right]\nonumber\\
&&\times\rho_{I,\{i_n,j_n\}}^Q(0),
\end{eqnarray}
where
\begin{equation}
c(\omega_k,t)=\frac{1-\cos(\omega_kt)}{{\omega_k}^2}.
\end{equation}

Taking the continuum limit, we get the form of equation (\ref{rhoi}), with
\begin{eqnarray}
\Theta(t)&=&\int d\omega I (\omega)s(\omega,t),
\end{eqnarray}
\begin{eqnarray}
\Gamma(t,T)&=&\int d\omega I (\omega) c(\omega,t)\coth\left(\frac{\omega}{2\omega_T}\right).
\end{eqnarray}
Here, $\omega_T\equiv k_BT$ is called the thermal frequency, and the spectral density
\begin{equation}
I(\omega)\equiv \sum_\mathbf{k}\delta(\omega-\omega_k)|g_\mathbf{k}|^2\equiv \frac{dk}{d\omega}G(\omega)|g(\omega)|^2,
\end{equation}
with $G(\omega)$ being the density of state.
Assuming an ohmic spectral density
\begin{equation}
I(\omega)=\eta\omega e^{-\omega/\omega_c},
\end{equation}
we get equation (\ref{ohmic1}) and (\ref{ohmic2}).

\pagebreak

\end{document}